\documentclass[]{aa}

\usepackage[T1]{fontenc}
\usepackage[utf8]{inputenc}

\usepackage{graphicx}
\usepackage{natbib}
\usepackage{ulem}
\usepackage{textcomp}
\usepackage{gensymb}
\usepackage{longtable}
\usepackage{threeparttable}
\usepackage{multicol}
\usepackage{multirow}
\usepackage{textgreek}
\usepackage{float}
\usepackage{todonotes}
\usepackage[Symbol]{upgreek}
\usepackage{amsmath}
\usepackage{etoolbox}
\usepackage{txfonts}
\usepackage{url}
\usepackage{xcolor}

\usepackage[most]{tcolorbox}
\usepackage{hyperref}
\usepackage{xcolor}
\hypersetup{
  colorlinks   = true, 
  urlcolor     = blue, 
  linkcolor    = blue, 
  citecolor   = blue 
}
\usepackage[all]{hypcap}

\begin{document} 
	\title{Deep low-frequency radio observations of Abell 2256 II: The ultra steep spectrum radio halo}
	\titlerunning{ The ultra steep spectrum radio halo in  Abell 2256}
	\authorrunning{Rajpurohit et al.}

\author{K. Rajpurohit\inst{1,2,3}, E. Osinga\inst{4}, M. Brienza\inst{5,1,2}, A. Botteon\inst{1,2,4}, G. Brunetti\inst{2}, W. R. Forman\inst{6}, C. J. Riseley\inst{1,2,7}, \\ F. Vazza\inst{1,2,8}, A. Bonafede \inst{1,2}, R. J. van Weeren \inst{4}, M. Br\"uggen\inst{8}, S. Rajpurohit\inst{9}, A. Drabent\inst{3}, D. Dallacasa\inst{1}, \\ M. Rossetti \inst{10}, A. S. Rajpurohit \inst{11}, M. Hoeft\inst{3}, E. Bonnassieux \inst{12}, R. Cassano\inst{2}, and G.K.Miley \inst{4}}

\institute{Dipartimento di Fisica e Astronomia, Universit\`a di Bologna, via P. Gobetti 93/2, 40129, Bologna, Italy\\
 {\email{kamlesh.rajpurohit@unibo.it}}
\and
INAF-Istituto di Radio Astronomia, Via Gobetti 101, 40129, Bologna, Italy
\and
Th\"uringer Landessternwarte (TLS), Sternwarte 5, 07778 Tautenburg, Germany
\and
Leiden Observatory, Leiden University, P.O. Box 9513, 2300 RA Leiden, The Netherlands
\and
INAF - Osservatorio di Astrofisica e Scienza dello Spazio di Bologna, via Gobetti 93/3, 40129 Bologna, Italy
\and
Harvard-Smithsonian Center for Astrophysics, 60 Garden Street, Cambridge, MA 02138, USA
\and
CSIRO Space \& Astronomy, PO Box 1130, Bentley, WA 6102, Australia
\and
Hamburger Sternwarte, Universit\"at Hamburg, Gojenbergsweg 112, 21029, Hamburg, Germany
\and
Molecular Foundry, Lawrence Berkeley National Laboratory, Berkeley, CA 94720, USA
\and
INAF - IASF Milano, via A. Corti 12, 20133 Milano, Italy
\and
Astronomy \& Astrophysics Division, Physical Research Laboratory, Ahmedabad 380009, India
\and
Lehrstuhl f\"ur Astronomie, Universit\"at W\"urzburg, Campus Hubland Nord, Emil-Fischer-Strasse 31, 97074 W\"urzburg, Germany
}

%############################################################################################################
% \abstract
%############################################################################################################   
  
%############################################################################################################
% \abstract
%############################################################################################################   
  
\abstract
{We present the first detailed analysis of the radio halo in the merging galaxy cluster Abell 2256 using the LOw Frequency ARray (LOFAR) and the upgraded Giant Metrewave Radio Telescope (uGMRT) and the VLA. Radio observations (120 MHz-2\,GHz) combined with archival \textit{Chandra} and \textit{XMM-Newton} X-ray data allowed us to study the halo emission with unprecedented detail. The integrated radio emission from the entire halo is characterized by an ultra steep spectrum, which can be described by a power law with $\alpha_{144\,\rm MHz}^{1.5\,\rm GHz}=-1.63\pm0.03$ with radial steepening in the outer regions. The halo is significantly underluminous according to the current scaling relations between radio power and mass at 1.4\,GHz but not at 150\,MHz; ultra steep spectrum halos are predicted to be statistically underluminous. Despite the complex structure of this system, the radio halo morphology is remarkably similar to that of the X-ray emission. The radio surface brightness distribution across the halo is strongly correlated with the X-ray brightness of the intracluster medium (ICM). The derived correlations show sublinear slopes and there are distinct structures:  the core is $\rm I_{R}\propto I_{X}^{1.51}$, the outermost region $\rm I_{R}\propto I_{X}^{0.41}$, and we find radio morphological connections with X-ray discontinuities. We also find a strong anti-correlation between the radio spectral index and the X-ray surface brightness, implying radial steepening. We suggests that the halo core is either related to old plasma from previous AGN activity, being advected, compressed and re-accelerated by mechanisms activated by the cold front or  less turbulent with strong magnetic field in the core. The change in the radio vs X-ray correlation slopes in the outer regions of the halo could be due to a radial decline of magnetic field, increase in the number density of seed particles or increasing turbulence. Our findings suggest that the emitting volume is not homogenous according to turbulent re-acceleration models.}


   \maketitle
%
%-------------------------------------------------------------------

%############################################################################################################
% introduction 
%############################################################################################################

\section{Introduction}
 \label{sec:intro}

Galaxy clusters undergoing mergers often show Megaparsec-scale radio relics and radio halos \citep[see][for a recent review]{Brunetti2014,vanWeeren2019}. These halos and relics provide direct evidence for the presence of magnetic fields and relativistic particles that are mixed with the thermal intracluster medium (ICM). The radio spectra{\footnote{$S_{\nu}\propto\nu^{\alpha}$, with spectral index $\alpha$}} of such sources are steep ($\alpha \leq-1.0$). The origin of the radiating relativistic particles that produce halos and relics has not been fully understood.

%####################################
%  Table 1: observing details 
%####################################
\setlength{\tabcolsep}{13pt}
\begin{table*}[!htbp]
\caption{Observational overview: uGMRT, LOFAR, and VLA observations.}
\centering
\begin{threeparttable} 
\begin{tabular}{ l  c  c c c c }
  \hline  \hline  
& LOFAR HBA$^{_\dagger}$& uGMRT Band\,3  &  uGMRT Band\,4   & VLA L-band $^{\ddagger}$  \\  
  \hline  
Frequency range &120-169 MHz&300-500\,MHz&550-950\,MHz&1-2\,GHz\\ 
Channel width & 12.2\,kHz & 97\,kHz &49\,kHz&1\,MHz \\ 
No of channels  &64 &4096 &4096&64\\ 
On source time &16\,hrs &10\,hrs &8\,hrs&24\,hrs \\
LAS$^{_\star}$ & 228\arcmin& 32\arcmin &$ 17\arcmin$& 16\arcmin\\ 
\hline 
\end{tabular}
\begin{tablenotes}[flushleft]
\footnotesize
\item{\textbf{Notes.}} Full Stokes polarization information was recorded for the uGMRT Band\,4, Band\,4,  and VLA L-band. For the VLA, the number of spectral window are 16 each with 64 channels; $^{_\dagger}$ for data reduction of the LOFAR 144\,MHz and uGMRT, VLA observations, we refer to Osinga et al. in prep. and paper I, respectively; $^{_\ddagger}$archival VLA L band data; $^{_\star}$Largest angular scale (LAS) that can be recovered by the mentioned observations. 
\end{tablenotes}
\end{threeparttable} 
\label{Tabel:obs}
\end{table*}

Radio halos are found at the centers of galaxy clusters where the radio emission typically follows the thermal X-ray surface brightness. One of the leading models involves the re-acceleration of cosmic-ray electrons (CRe) by the interaction with turbulence injected into the ICM in connection with cluster mergers \citep{Brunetti2001,Petrosian2001,Brunetti2007,Beresnyak2013,Miniati2015,Brunetti2016}. This model can explain the connection with cluster dynamics \citep{Cassano2010,Cassano2013,Cuciti2021} and the complex spectral properties observed in the population of radio halos, including the existence of ultra steep spectrum ($\alpha \leq-1.5$) halos \citep{Brunetti2008,Dallacasa2009,Wilber2018,Rajpurohit2021b, Luca2021,Gennaro2021,Duchesne2021} and the presence of large-scale fluctuations in the spectral index distribution \citep{Botteon2020a,Rajpurohit2021b,Rajpurohit2021c,Bonafede2022}. 

Gamma ray limits for clusters hosting radio halos suggest that secondary emission from p-p collisions between relativistic and thermal protons in the ICM are subdominant \citep{Brunetti2012,Ackermann2014,Ackermann2016}, yet turbulent reacceleration of these secondary particles is still consistent with radio and gamma-ray observations \citep[][]{Brunetti2017,Pinzke2017,Adam2021}. In recent years, the picture has become more complicated as some clusters hosting mini halos, believed to be associated with sloshing motions of the ICM, have been found to also contain a large-scale steep spectrum radio component with properties similar to radio halos. \cite[e.g.,][]{Biava2021,Chris2022}.

In this paper, we focus on the radio halo in the galaxy cluster Abell 2256. A detailed analysis of the large filamentary relic is presented in \cite{Rajpurohitpaper1} (hereafter paper I). Due to strong emission at radio and X-ray wavelengths, low redshift, the presence of several X-ray surface brightness discontinuities (cold fronts and shock fronts), and its large angular extent, the halo in Abell 2256 is excellently suited to test currently proposed acceleration models for radio halo formation. High-resolution radio observations over a wide frequency range are essential to improve our understanding of the particle acceleration in radio halos. Since halos are low surface brightness sources, only a few systems, including Abell 2256, allow for detailed investigations with current instruments. We use data from the LOw-Frequency ARray (LOFAR), upgraded Giant Metrewave Radio Telescope (uGMRT), and the Karl G. Jansky Very Large Array (VLA). The uGMRT observations are originally published in paper I and LOFAR high band antenna (HBA) in Osinga et al. (in prep.). These sensitive wideband observations allow us to study the low-surface brightness halo emission at a higher resolution than has been done previously. To compare the radio and X-ray properties and the connection between the thermal and nonthermal plasma of the ICM, we used archival \textit{Chandra} and \text{XMM-Newton} observations.

The layout of this paper is as follows: In Sect.\,\ref{observations}, we present an overview of the observations and data reduction. The new radio images are presented in Sect.\,\ref{results}. The results obtained with radio and X-ray analysis are described in Sect.\,\ref{halo_anaylsis} followed by a summary in Sect.\,\ref{summary}.

Throughout this paper, we adopt a flat $\Lambda$CDM cosmology with $H_{\rm{ 0}}=69.6$ km s$^{-1}$\,Mpc$^{-1}$, $\Omega_{\rm{ m}}=0.286$, and $\Omega_{\Lambda}=0.714$. At the cluster's redshift, $1\arcsec$ corresponds to a physical scale of 1.13\,kpc. All output images are in the J2000 coordinate system and are corrected for primary beam attenuation. 

%%%%%%%%%%%%%%%%%%%%%%%%%%%
%Figure 1
%%%%%%%%%%%%%%%%%%%%%%%%%%%
 
  \begin{figure*}[!thbp]
    \centering
    \includegraphics[width=1\textwidth]{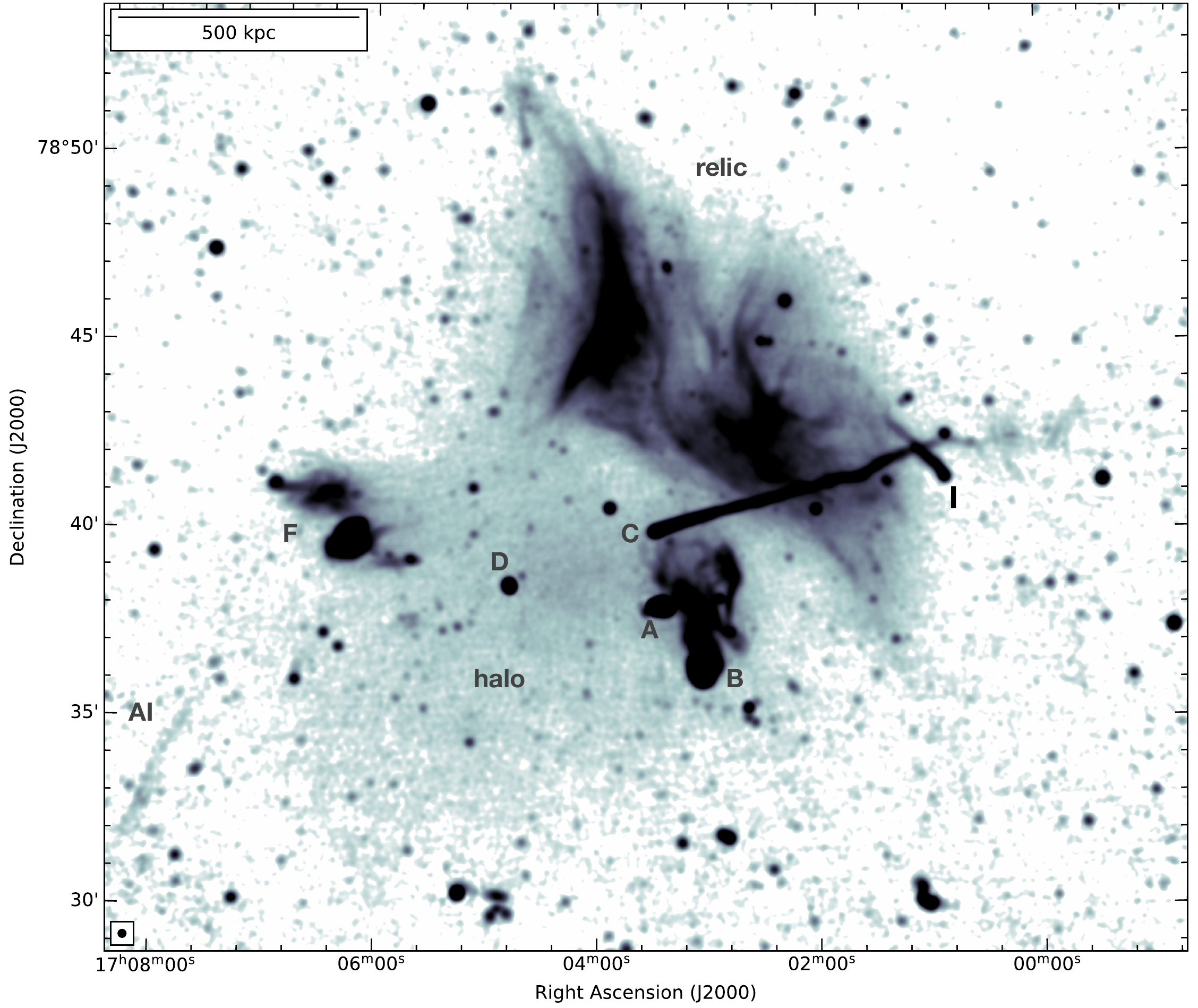}
 \caption{uGMRT Band\,4 (550-850\,MHz) image of the galaxy cluster Abell 2256 with an angular resolution of  $12\arcsec$, highlighting the spectacular central halo emission, the filamentary relic, and radio galaxies in the field. The beam size is indicated in the bottom left corner of each image. The image is created with Briggs weighting using ${\tt robust}=-0.5$. The rms is $\rm 14 \mu Jy\,beam^{-1}$.}
      \label{labelling_halo}
  \end{figure*}    

%########################################################################
% Section 2 Previous studies 
%########################################################################

\section{Abell 2256}
\label{target}
Abell 2256 is a massive nearby cluster ($z=0.058$) that exhibits strong emission at all wavelengths. In the radio domain (see Figs.\,\ref{labelling_halo} and \ref{low_res}), the cluster is characterized by complex diffuse emission on a large scale that consists of an extended steep spectrum radio halo at the center of the cluster, which is surrounded by a giant filamentary relic to the northwest and several complex radio sources \citep{Bridle1976,Bridle1979,Rottgering1994,Kim1999,Clarke2006,Brentjens2008,vanWeeren2009,vanWeeren2012b,Owen2014,Rajpurohitpaper1}. The existence of a radio halo in Abell 2256 was first suggested by \cite{Bridle1979} using WSRT observations at 610\,MHz. The presence of halo emission was shown at both low- and high-frequency from 63\,MHz to 1.4\,GHz \citep{Clarke2006,Brentjens2008,Kale2010,vanWeeren2012b,Owen2014}. In fact, some of these studies reported the properties of the large relic, which was misinterpreted as halo emission. Therefore, detailed radio morphology, size, and characterization of the halo emission remained missing, mostly because of poor resolution, the lack of adequate sensitivity, and possible overlap between the halo, relic, and nearby radio galaxies.

The cluster has a total mass of about $\rm M_{500}=(6.1\pm0.4)\times10^{14}\,M_{\odot}$ \citep{Planck2011,Markevitch1997}. The dynamical state of the cluster is complex, as suggested by both optical and X-ray observations. Optical studies of the galaxy distribution reveal that the cluster consists of three separate mass components \citep{Berrington2002,Miller2003}. These studies provide strong evidence that Abell 2256 is undergoing a merger event between a main cluster component, a major subcomponent, and a third infalling group \citep{Fabricant1989,Berrington2002,Miller2003,Ge2020}.  

At X-ray wavelengths, the cluster is luminous \citep[$L_{\rm X,0.1-2.4\,\rm keV}=3.7\times 10^{44}\,\rm erg\,s^{-1}$;][]{Ebeling1998}. The cluster is very rich in X-rays consisting of several substructures, \citep{Fabricant1989,Briel1991,Briel1994,Roettiger1995,Molendi2000,Sun2002,Ge2020,Breuer2020}. \textit{Chandra} and \textit{XMM-Newton} studies provide evidence for several X-ray surface brightness discontinuities \citep{Sun2002,Bourdin2008,Trasatti2015,Ge2020,Breuer2020}.

%########################################################################
% Chandra data
%########################################################################

\setlength{\tabcolsep}{4pt}
\begin{table}[!htbp]
    \caption{\textit{Chandra} and \textit{XMM-Newton} observations used in this work. The net time for \textit{XMM-Newton} refers to the net time for each detector (mos1,mos2,pn).}
    \centering
    \begin{tabular}{l l l l}
    \hline
   & ObsID & Net time (ks) & Observing date \\
    \hline
    \hline
      &16129 & 44.5 & August, 14 2014\\
    &16514 & 44.5 &August, 17 2014\\
\textit{Chandra}   &16515 & 43.2 &September, 07 2014\\
    &16516 & 44.5 &September, 26 2014\\
    \hline
     & 0112500201 & 10.1, 8.6, 4.2 & March, 20 2002\\
  &0112950501 & 3.8, 3.1, 0.6 &March, 20 2002\\
    &0112950601 & 10.8, 11.8, 5.9 &March, 28 2002\\
\textit{XMM-} & 0112950801 & 11.0, 5.4, 0.8 &March, 30 2002\\ 
\textit{Newton} &  0112950901& 0.5, 0.5, 1.7 &April, 23 2002\\ 
   &  0401610101& 38.9, 35.3, 14.5 &April, 19 2002\\ 
    &  0112951501& 8.5, 8.9, 6.4 &June, 02 2002\\ 
     &  0112951601& 10.3, 10.8, 5.5 &September, 22 2002\\ 
    \hline    
    \end{tabular}
    \label{tab:chandra}
\end{table}

%########################################################################
% Section 3: Observations and data reduction
%########################################################################

\section{Observations and data reduction}
\label{observations}
\subsection{Radio observations: LOFAR, uGMRT, and VLA}
We observed the cluster with the upgraded GMRT in Band\,4 (550-950\,MHz) and Band\,3 (300-450\,MHz), and LOFAR (120-168\,MHz; Osinga et al. in prep). We also used archival VLA observations. In Table\,\ref{Tabel:obs}, we summarize the observational details. For a detailed description of uGMRT, LOFAR, and VLA observations and the data reduction procedure, we refer to Paper I. In summary, the wideband uGMRT data were processed using the Source Peeling and Atmospheric Modeling \citep[$\tt{SPAM}$;][]{Intema2009}, pipeline{\footnote{\url{http://www.intema.nl/doku.php?id=huibintemaspampipeline}}}. The LOFAR HBA data reduction and calibration were performed with the LoTSS DR2 pipeline \citep{Tasse2020} followed by the final ``extraction+self-calibration'' scheme \citep{vanWeeren2020}. The VLA A, B, C, and D configuration data were calibrated in {\tt CASA} (paper I). 

The final deconvolution of all data was performed in {\tt WSClean} \citep{Offringa2014} using $\tt{multiscale}$ and $\tt{Briggs}$ weighting with a robust parameter $-0.5$. The output images presented in this work were corrected for primary beam attenuation. The uncertainty in the flux density measurements was estimated as:
\begin{equation}
\Delta S =  \sqrt {(f \cdot S)^{2}+{N}_{{\rm{ beams}}}\ (\sigma_{{\rm{rms}}})^{2}},
\end{equation}
where $f$ is an absolute flux density calibration uncertainty, $S_\nu$ is the flux density, $\sigma_{{\rm{ rms}}}$ is RMS noise, and $N_{{\rm{beams}}}$ is the number of beams. We assume absolute flux density uncertainties of 10\,\% for uGMRT Band\,3 \citep{Chandra2004} and LOFAR data, \citep{Shimwell2022} 5\% for uGMRT Band\,4 data, and 2.5\,\% for the VLA data \citep{Perley2013}.

\subsection{X-ray observations: Chandra and XMM-Newton}
We reprocessed the five archival \textit{Chandra} ACIS-I observations of Abell 2256 that are listed in Table~\ref{tab:chandra} together with their individual net exposure time. We note that three other ACIS-S observations are present in the Chandra archive (ObsIDs: 965, 1521, 2419) but these were not considered in this work because of the smaller field of view of ACIS-S and shorter exposure times. Data were processed with {\tt CIAO} v4.12 and \textit{Chandra} {\tt CalDB} v4.9.0 using \texttt{chandra\_repro} to produce new level=2 event files. Periods of observations affected by soft proton flares were removed with task \texttt{deflare} after inspecting the light curves extracted from the S2 chip in the 0.5$-$0.7 keV band. The five ObsIDs were combined with \texttt{merge\_obs} to produce an exposure-corrected mosaic image of the cluster in the 0.5$-$2.0 keV band. 

We combined the individual exposure-corrected point spread function (PSF) of each ObsID to produce a single PSF map with minimum size that was passed to the task \texttt{wavdetect} to detect point sources in the mosaiced image. Point sources were detected using wavelet radii of 1, 2, 4, 8, and 16 pixels that are removed by replacing the pixels inside the point source regions with randomized values from surrounding regions.

We also produced an \textit{XMM-Newton} mosaic of Abell 2256 in the 0.5-2.0 keV band that was obtained using 8 archival observations (for ObsIDs, see Table\,\ref{tab:chandra}):. Each single ObsID was processed using the XMM-Newton Scientific Analysis System ({\tt SAS v16.1.0}) and the Extended Source Analysis Software (ESAS) data reduction scheme. Then, count images, background images, and exposure maps of each observation were combined to produce background-subtracted and exposure-corrected image. This image covers a larger field of view than \textit{Chandra} observations and was used to compare the radio and X-ray surface brightness of the ICM.

%%%%%%%%%%%%%%%%%%%%%%%%%%%
%Figure 2
%%%%%%%%%%%%%%%%%%%%%%%%%%%

  \begin{figure*}[!thbp]
    \centering
    \includegraphics[width=1\textwidth]{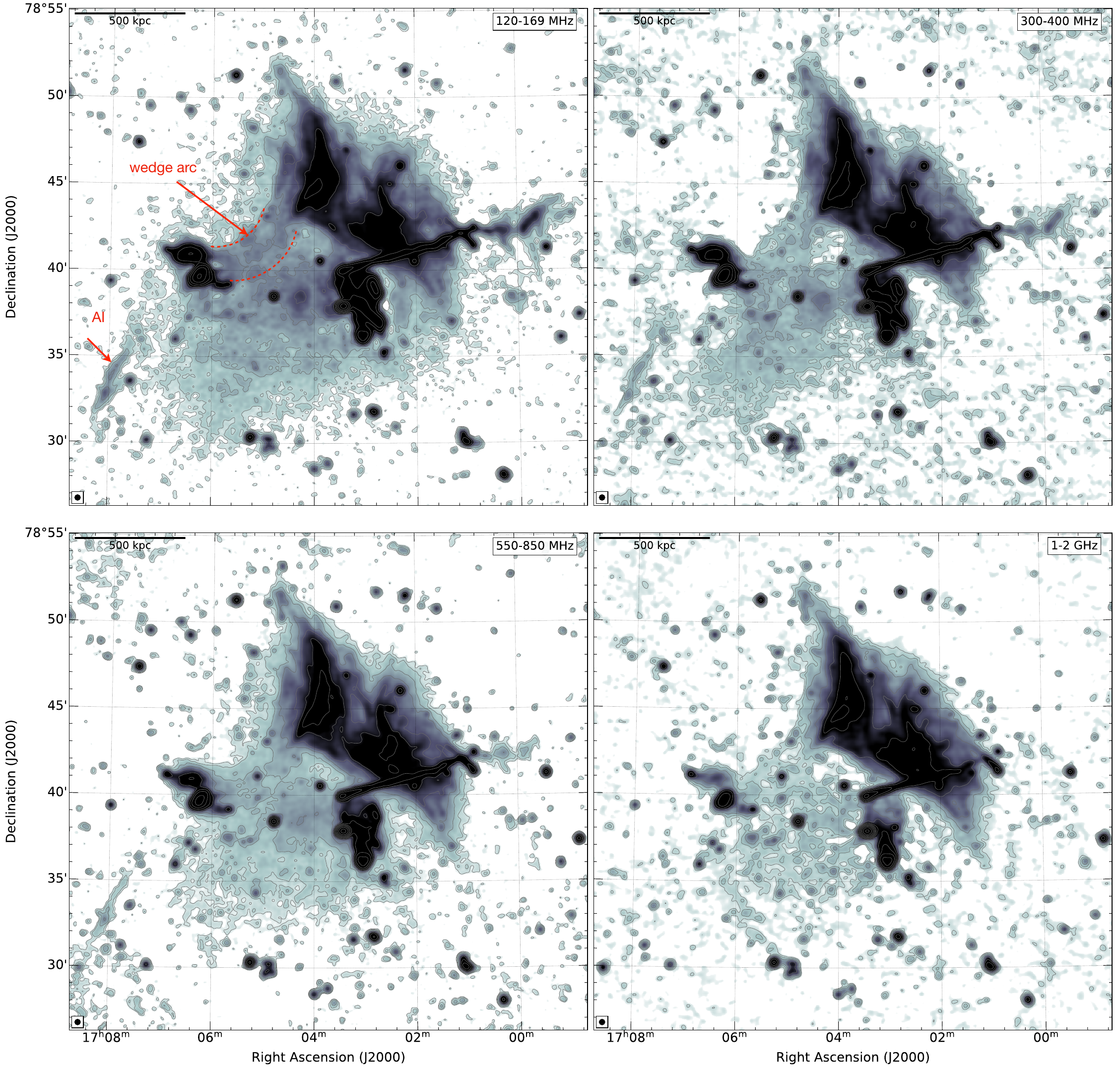}
 \caption{Low resolution LOFAR 144\,MHz (Osinga et al. in prep), uGMRT (300-950\,MHz), and VLA L-band images of Abell 2256 in square root scale. All images are created at a common $20\arcsec$ resolution. The radio surface brightness unit is $\rm mJy\,beam^{-1}$. The beam size is indicated in the bottom left corner of each image. Contour levels are drawn at $[1,2,4,8,\dots]\,\times\,3.0\,\sigma_{{\rm{ rms}}}$, where $\rm \sigma_{144\,MHz,\,rms}=180\mu Jy\,beam^{-1}$, $\rm \sigma_{350\,MHz,\, rms}=44\mu Jy\,beam^{-1}$, $\rm \sigma_{675\,MHz, \,rms}=18\mu Jy\,beam^{-1}$ and $\rm \sigma_{1.5\,GHz,\,rms}=10\mu Jy\,beam^{-1}$. These images highlights the morphology of the halo emission with observing frequency and show that the halo is much more extended toward low frequencies.  There are two other distinct features: an arc-shaped wedge to the east (show within red dotted curves) and source AI to the south-west }
      \label{low_res}
  \end{figure*}

  %############################################################################################################   
  \section{Results: Radio and X-ray morphology}
\label{results}
  %############################################################################################################   

%%%%%%%%%%%%%%%%%%%%%%%%%%%
%Figure 3
%%%%%%%%%%%%%%%%%%%%%%%%%%%

\begin{figure*}[!thbp]
    \centering
        \includegraphics[width=1.00\textwidth]{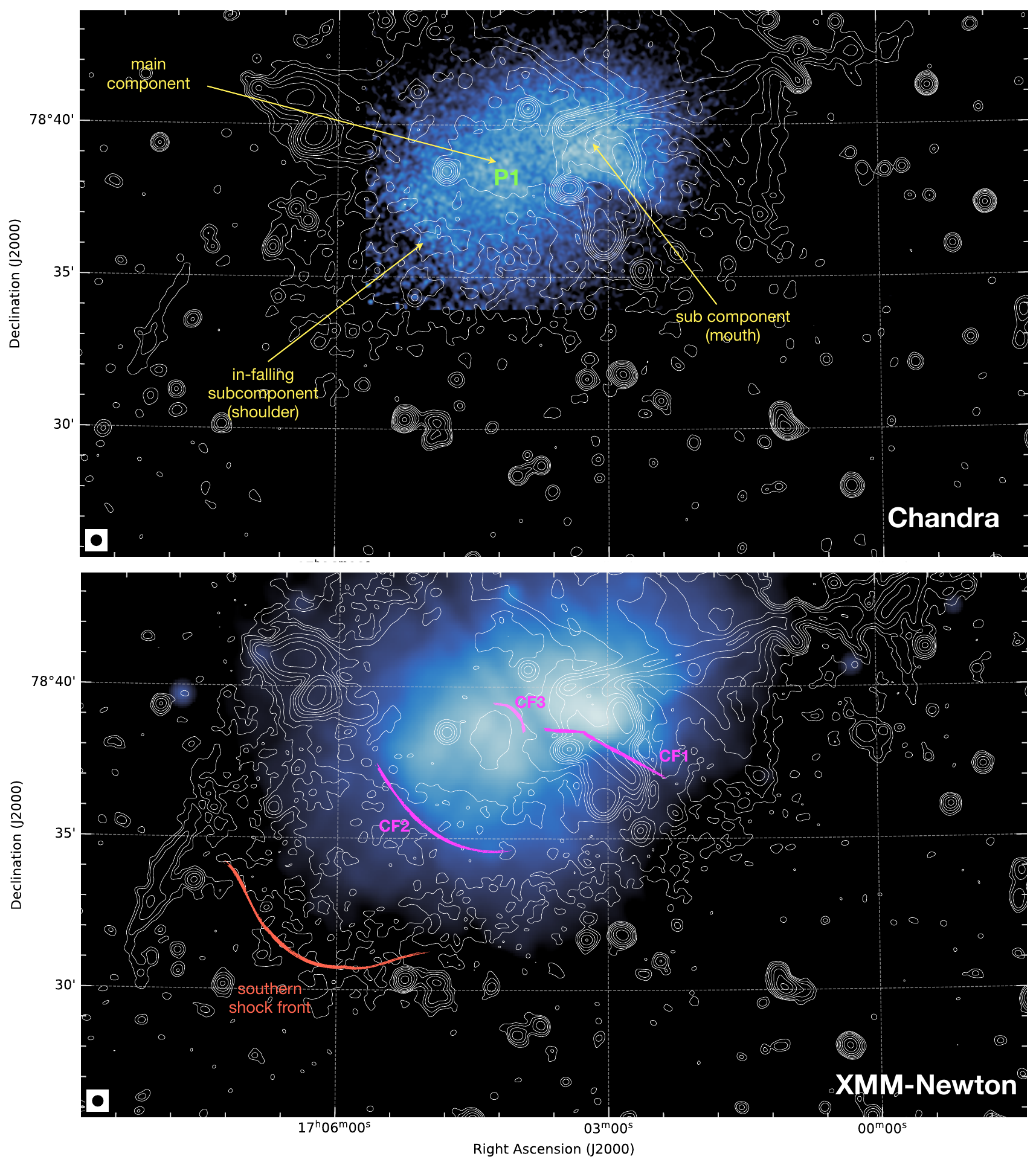}      
        \vspace{-0.6cm} 
 \caption{\textit{Top}: \textit{Chandra} $0.5-2.0$\,keV band point source subtracted image, smoothed with a Gaussian full width at half maximum (FWHM) of 3\arcsec, overlaid with uGMRT Band\,4 radio contours at 20\arcsec resolution. \textit{Bottom}: \textit{XMM-Newton} image overlaid with LOFAR 144\,MHz radio contours at $20\arcsec$ resolution. Contour levels are drawn at $[1,2,4,8,\dots]\,\times\,4\,\sigma_{{\rm{ rms}}}$. The radio beam size is indicated in bottom left corner of the image. The labelling of cold fronts are done follwing \cite{Ge2020}. The overlay shows that the radio halo morphology nicely follows the thermal X-ray emission morphology.} 
\label{halo_Xray}
\end{figure*}  

%%%%%%%%%%%%%%%%%%%%%%%%%%%
%Figure 4
%%%%%%%%%%%%%%%%%%%%%%%%%%%

\begin{figure*}[!thbp]
    \centering
        \includegraphics[width=0.95\textwidth]{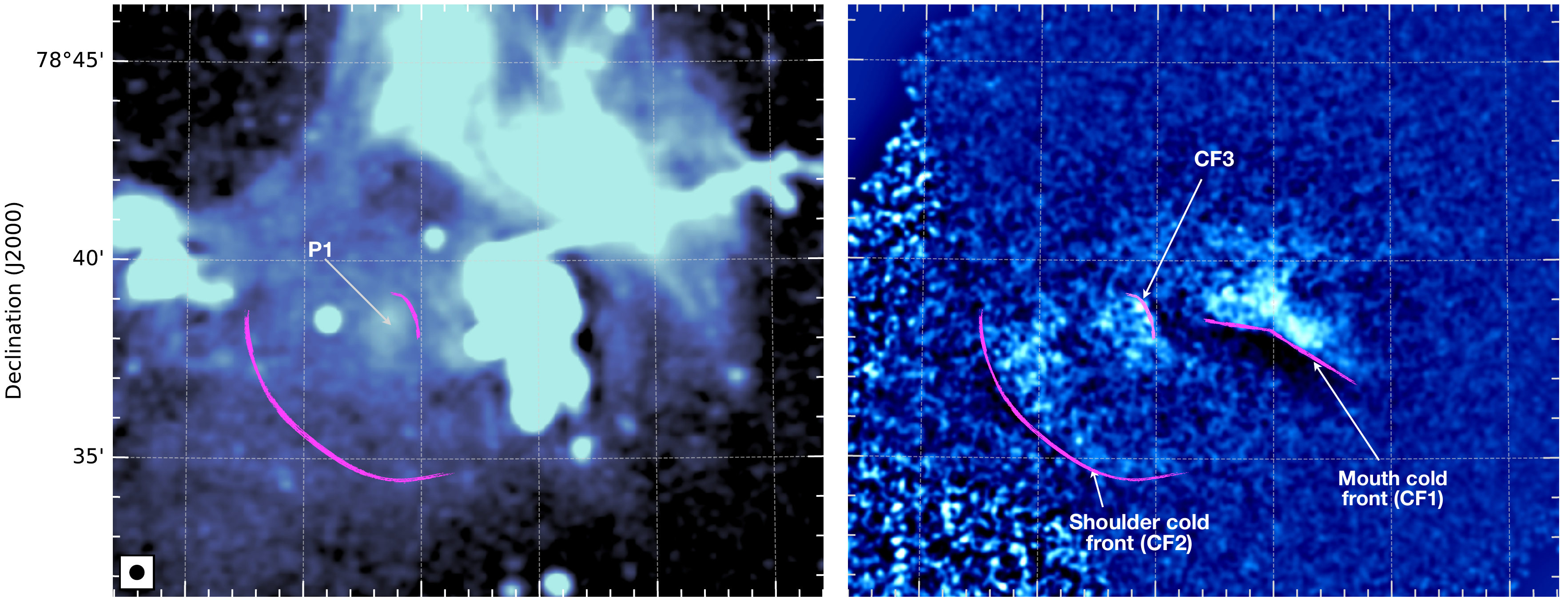}    
                \includegraphics[width=0.95\textwidth]{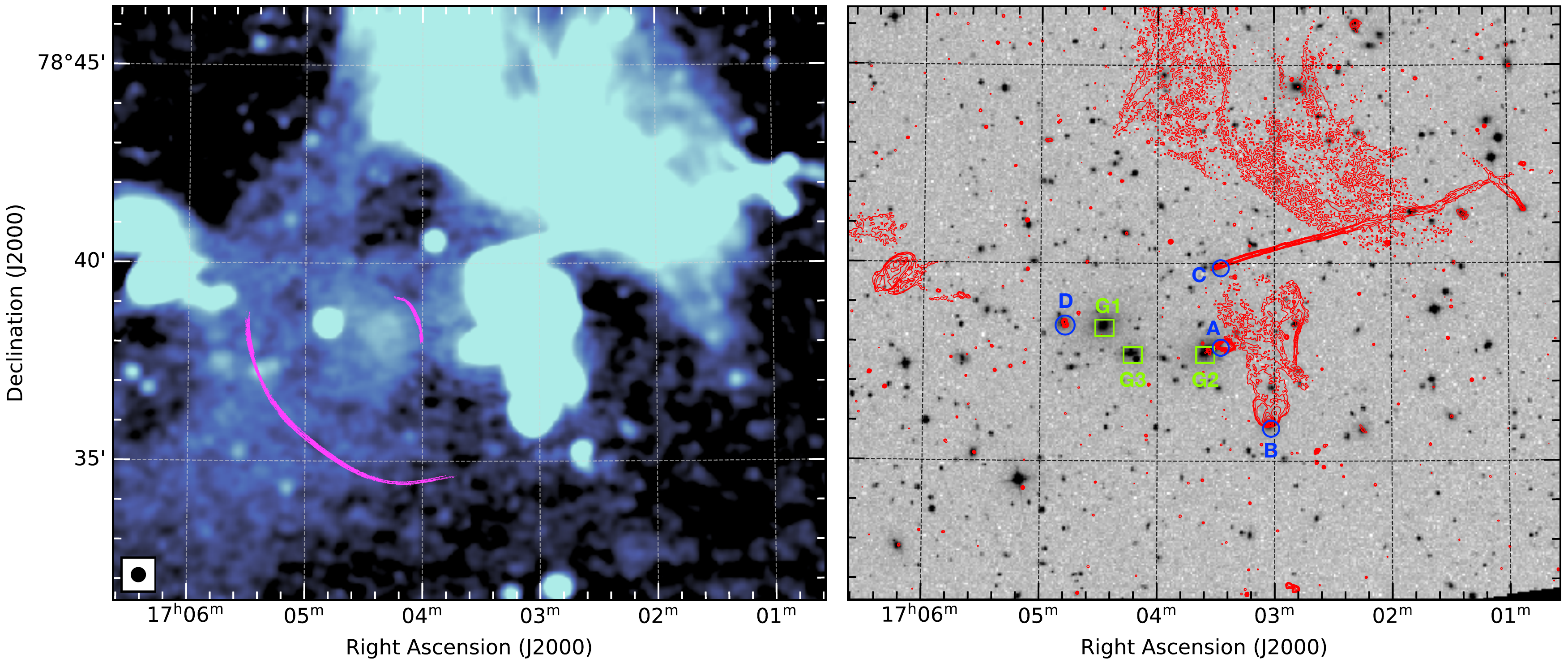}   
 \caption{\textit{Left}: LOFAR 144\,MHz (top) and GMRT 350\,MHz (bottom) image at 20 \arcsec resolution. To compare radio morphologies at these two frequencies, the colors in both images were scaled manually. \textit{Right}: Unsharp-masked \textit{Chandra} $0.5-2.0$\,keV image of the cluster A2256 created by subtracting images convolved with Gaussians with $\sigma_1$ (4\arcsec) and $\sigma_2$ (40\arcsec) and dividing by the sum of the two (top). The image displays the sharp edges in the X-ray surface brightness image.  The labeling of cold fronts (mouth and shoulder) is done following \cite{Sun2002}. The comparison of both images reveal that the peak P1 in the radio surface brightness within the main component is confined by the innermost cold front CF1. DSS image overlaid with a high resolution 675\,MHz radio contours (bottom). Green square boxes and labels mark the bright cluster galaxies in the halo. Blue circles and labels denote the position of known radio galaxies.  } 
\label{unsharp}
\end{figure*}

%%%%%%%%%%%%%%%%%%%%%%%%%%%
%Figure 5
%%%%%%%%%%%%%%%%%%%%%%%%%%%

 \begin{figure*}[!thbp]
    \centering
        \includegraphics[width=0.48\textwidth]{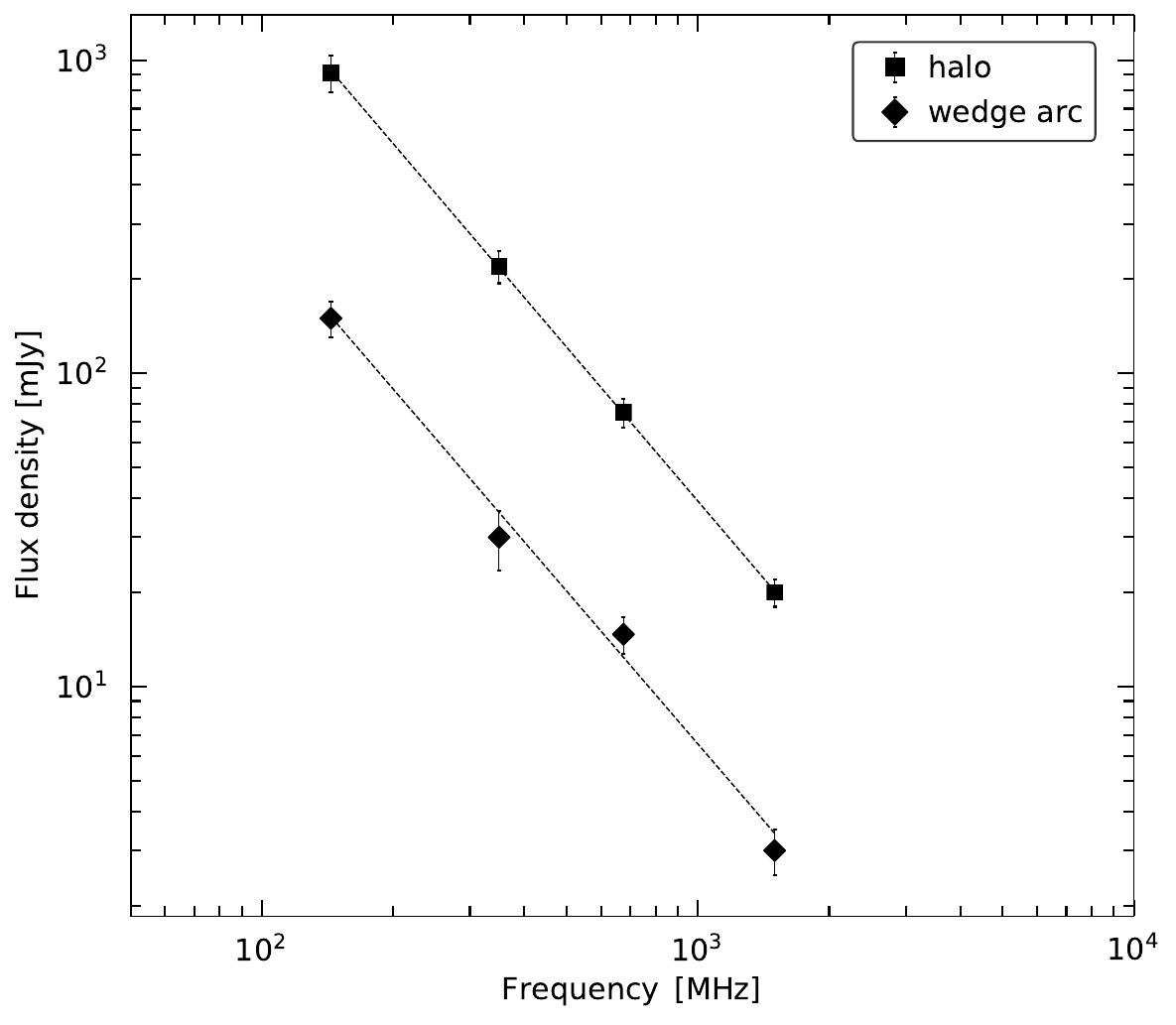}
         \includegraphics[width=0.47\textwidth]{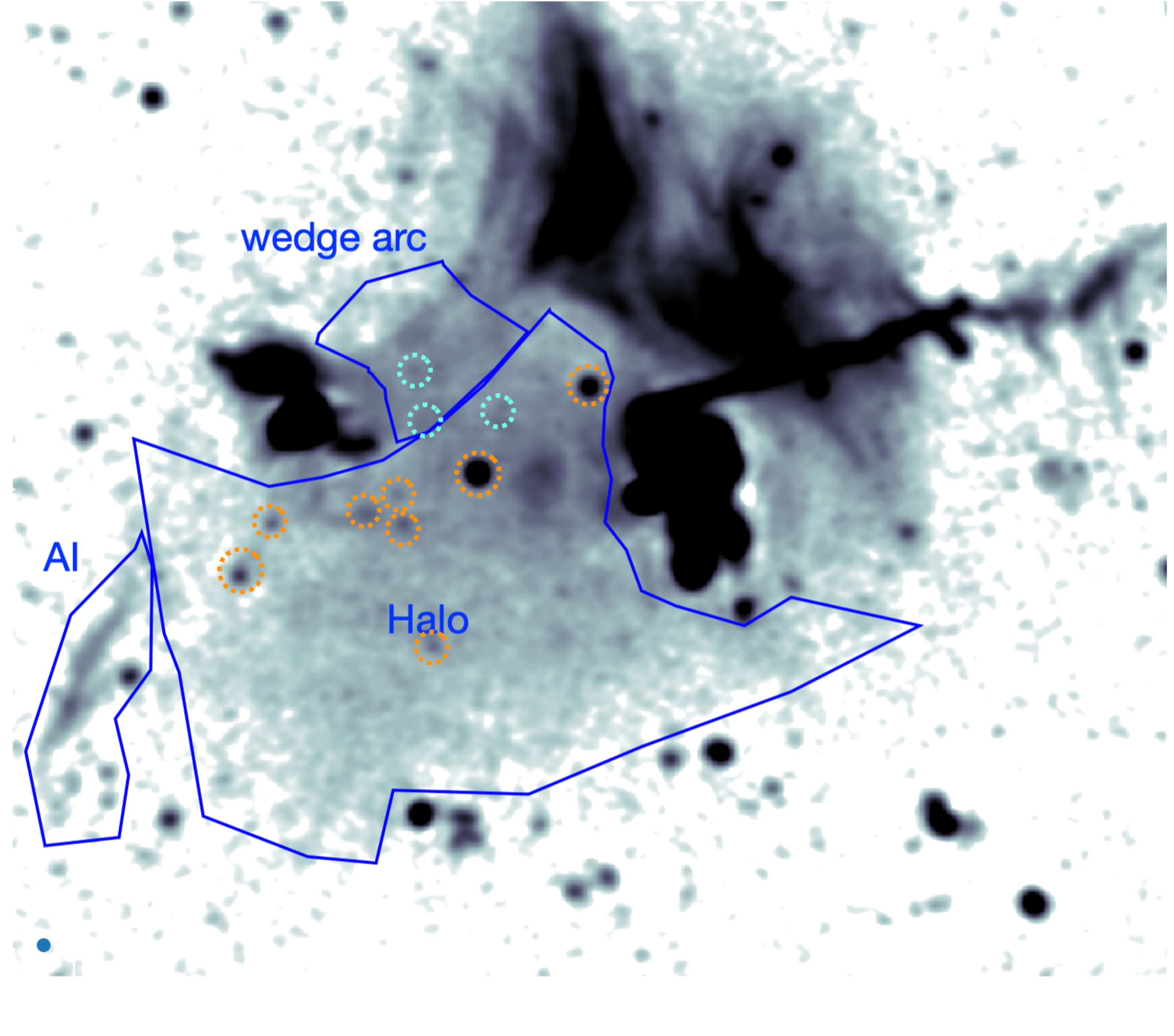}
         \includegraphics[width=0.48\textwidth]{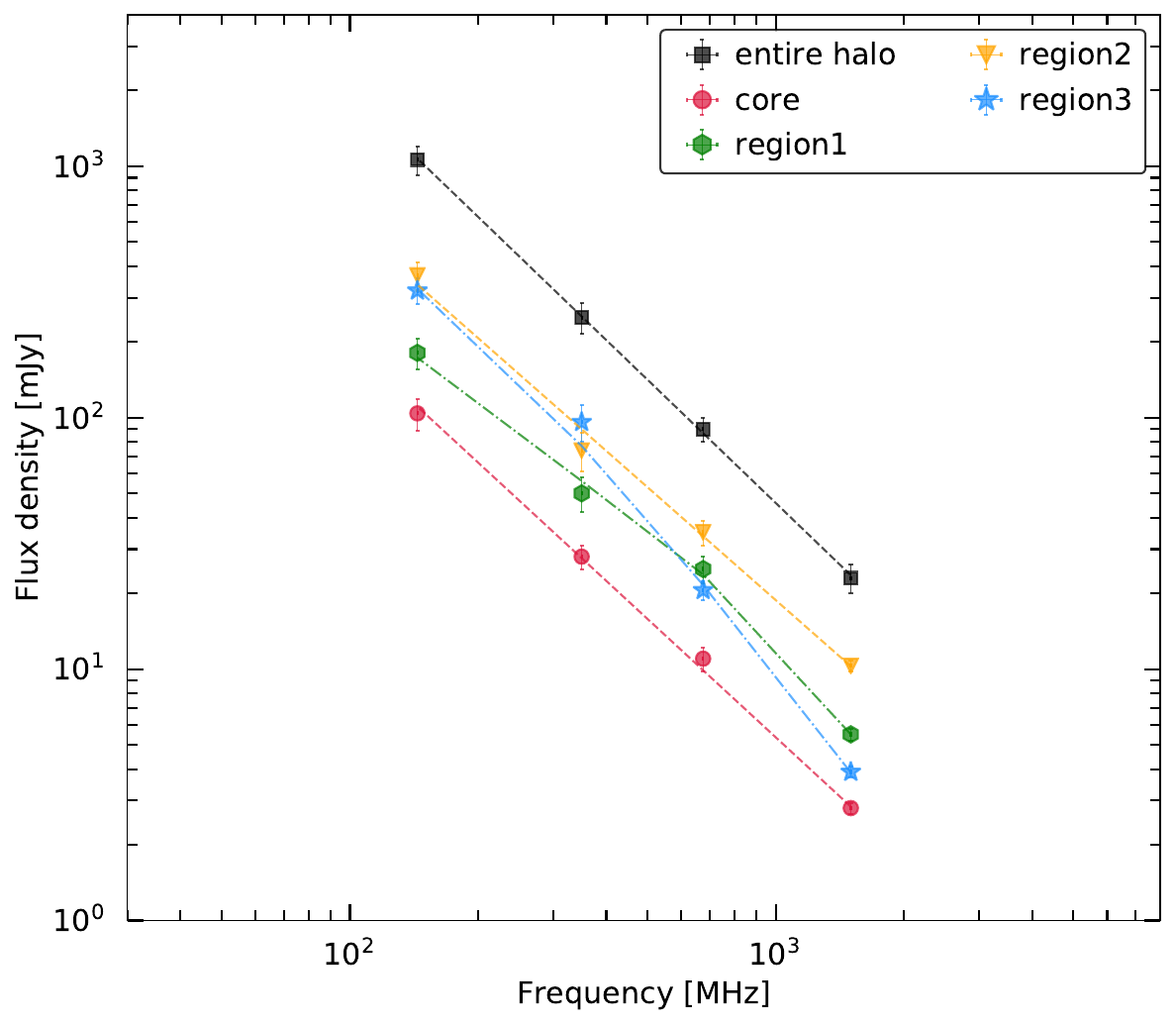} 
        \includegraphics[width=0.46\textwidth]{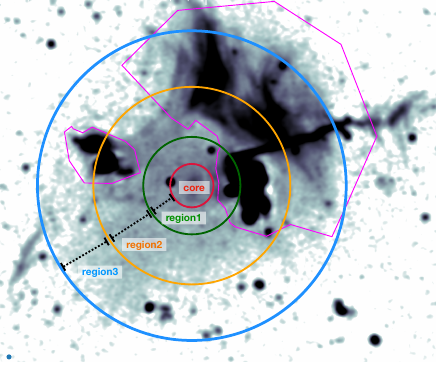}
 \caption{ \textit{Top left}: Integrated spectrum of the radio halo and the wedge arc between 144\,MHz and 1.5\,GHz. Dashed lines show the fitted power law. The overall spectrum of the halo follows a single power law and has an ultra steep spectral index of $-1.63\pm0.03$. \textit{Bottom left}: The halo subregions spectra, showing different spectral indices. Dashed lines show the fitted power law and dot-dashed fitted with curved spectra.  \textit{Top right}: LOFAR 144\, MHz image overlaid with regions used for extracting flux densities of the halo, wedged arc and source AI. Flux density contributions from compact sources (shown with dashed circles) were manually subtracted from the total halo flux density. Cyan circles denote compact sources that are detected in the uGMRT and VLA images. \textit{Bottom right}: The halo subregions: core is the innermost part and region3 the outermost one. These regions are used for extracting the flux densities and the radio and X-ray surface brightness in the halo subregions.  Magenta polygons depict masked regions, i.e., not included for measuring the halo flux densities.}
\label{halo_index}
\end{figure*}

%############################################################################################################   
\subsection{Radio halo emission}
%############################################################################################################   
The uGMRT 550-850\,MHz image of the cluster Abell 2556 is shown in Fig.\,\ref{labelling_halo} at a resolution of $12\arcsec$. The image shows a large filamentary relic to the north and several peculiar radio galaxies, see paper I. The central region of the cluster is dominated by the radio galaxies A, B, and C, and large-scale low surface brightness halo emission, which is the primary focus of this paper. 

In comparison to all the previously published total power continuum images, our detection of the halo is of higher significance and the morphology of the halo is more precisely characterized. As shown in Fig.\,\ref{labelling_halo}, at 550-850\,MHz, the large part of the halo is apparently elongated in the north-west and south-east direction, i.e., along the merger direction. \cite{Brentjens2008} reported the presence of filamentary sub-structures within the halo region. In contrast, none of our new images show filamentary sub-structures embedded within the halo region. The total extent of the halo emission is larger than previously reported  at a similar frequency \citep{Bridle1976,Intema2009}. In the published 610 MHz images, mainly the innermost part of the halo is detected, while the low surface brightness emission is not completely recovered very likely because of the poor uv-coverage at short baselines and low sensitivity. From our new image, the exact largest linear size (LLS) of the halo is difficult to measure because of the presence of the relic and several other complex sources. But the LLS of the halo is at least $\sim$ 750\,kpc at 675\,MHz. Additionally, we found at least 48  unrelated compact sources ($>3\sigma_{\rm rms}$) embedded in the halo region at 675\,MHz.

For morphological comparison, in Fig.\,\ref{low_res} we show the moderate resolution ($20\arcsec$) LOFAR (120-169\,MHz), Band\,3 (300-400\,MHz), Band\,4 (550-850\,MHz), and VLA (1-2 GHz) images. The shape of the radio halo is similar in these images. From Fig.\,\ref{low_res}, it is evident that the halo is more extended toward low frequencies. We emphasize that the radio observations presented in this paper are very deep and sensitive to low surface brightness emission. The halo appears to shrink at high frequencies due to the steepening in the outermost regions. This has been recently observed also in other giant halos, for example MACS\,J0717+35 and Abell 2744 \citep{Rajpurohit2021b, Rajpurohit2021c}.  The LLS size of halo at 144\,MHz  and 1.5\,GHz is  900\,kpc and 480\, kpc, respectively. The entire halo covers an area of about $\rm 721\,kpc \times 900\,kpc$, $\rm 560\,kpc \times 750\,kpc$, $\rm 490\,kpc\times 480\,kpc$ at 144\,MHz, 675\,MHz, and 1.5\,GHz, respectively. 

In 144 and 350\,MHz images, the radio surface brightness is enhanced in the region between source D and source A. The dominant blob-like emission feature to the west of source D coincides with the cluster core; see Fig.\,\ref{low_res}. The radio surface brightness across the halo is fainter than the relic emission and sources F and B. In radio maps, in particular at 144\,MHz, the southern edge of the halo follows a V-shaped morphology with the bottom of the V at the south-east.

A distinct morphological structure is an arc-shaped wedge (see Fig.\,\ref{low_res}) at the north east of the cluster, first reported by \cite{Owen2014} at 1-2\,GHz. No feature is detected in {\it Chandra} or {\it XMM-Newton} X-ray images \citep{Ge2020} at this location. The arc-shaped feature is apparently connected to the large relic, the central halo emission, and the source F. On the basis of only the visual appearance, whether the connection between these features is real or just in projection is not obvious. We note that the wedge arc is very likely associated with the halo, as discussed later in Sect.\,\ref{radio-versus-xray}, therefore in the remainder of the paper this region is considered as part of the halo emission.

There is a linear structure, labeled AI in Fig.\,\ref{low_res}, to the southeast of the cluster. It is located at a projected distance of about 1~Mpc from the cluster center and was previously detected by \cite{vanWeeren2009a} and \cite{Intema2009}.  We do not find any optical counterpart for this source. At 350\,MHz, the LAS of the source is 403\arcsec corresponding to a physical size of 455\,kpc at the cluster redshift. We measure a flux density of $4.3\pm0.2\,\rm mJy$ at 675\,MHz. The source is not detected at 1-2\,GHz and its overall spectrum is curved. We obtained a radio power of $P_{\rm 1.5 \,GHz}=3.7\times10^{22}\,\rm W\,Hz^{-1}$ using the spectral index of $-1.9$ between 350 and 675\,MHz. 

%%%%%%%%%%%%%%%%%%%%%%%%%%%
%Figure 6
%%%%%%%%%%%%%%%%%%%%%%%%%%%

 \begin{figure*}[!thbp]
\centering
   \includegraphics[width=0.7\textwidth]{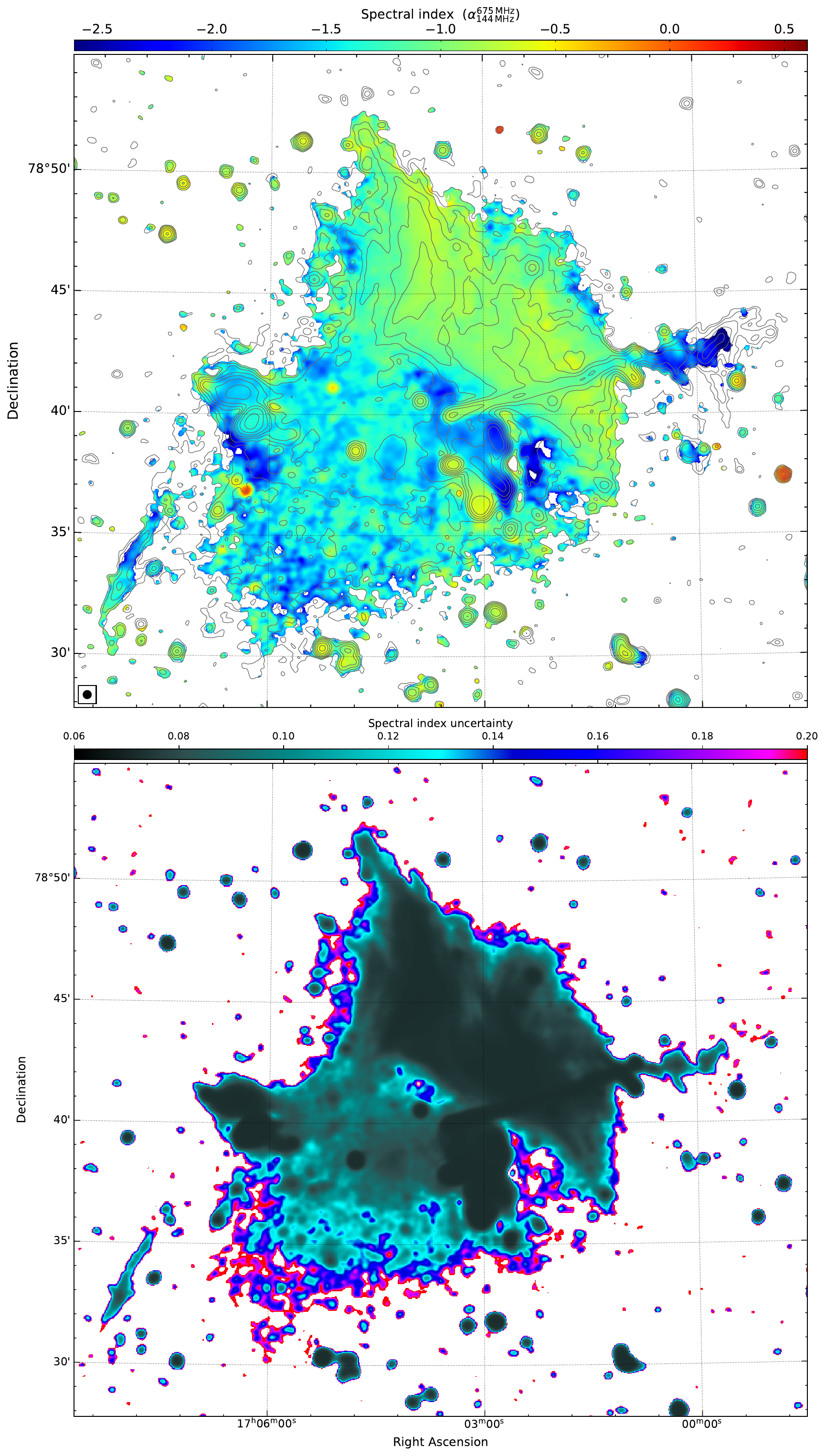}    
 \caption{\textit{Top}:  Spectral index map of the halo at 20\arcsec resolution between 144 and 675\,MHz. The halo spectral index varies on the scale of about 20\,kpc. Contour levels are drawn at $[1,2,4,8,\dots]\,\times\,3.0\,\sigma_{{\rm{ rms}}}$ and are from the LOFAR 144\,MHz image. \textit{Bottom}: Corresponding spectral index uncertainty. The beam size is indicated in the bottom left corner of the image.}
\label{halo_spectral_index_map}
\end{figure*}     

%%%%%%%%%%%%%%%%%%%%%%%%%%%
%Figure 7
%%%%%%%%%%%%%%%%%%%%%%%%%%%

 \begin{figure}[!thbp]
    \centering
        \includegraphics[width=0.49\textwidth]{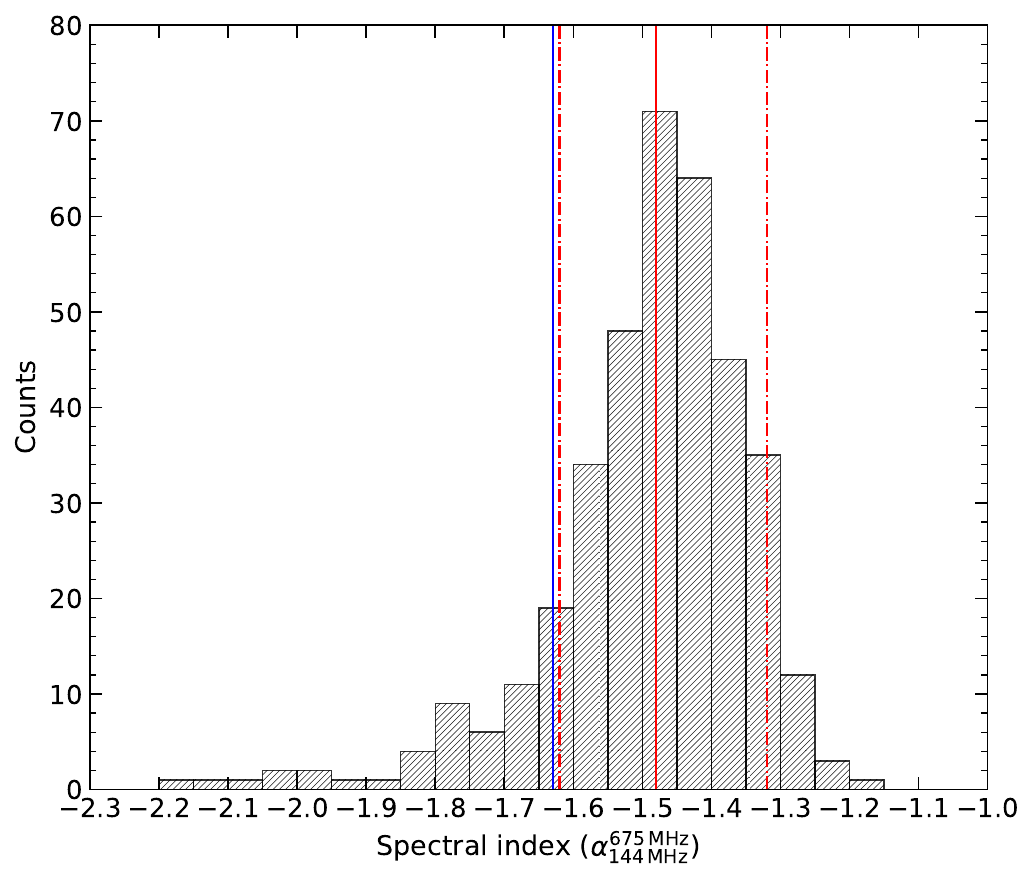}
 \caption{Histogram of the spectral index distribution across the halo in Abell 2256 between 144\,MHz and 675\,MHz. The spectral indices were extracted from square-shaped boxes with width $20\arcsec$, corresponding to a physical size of about 23\,kpc. The solid red line represents the median spectral index, namely $\langle\alpha\rangle-1.48$. The dash dotted lines show the standard deviation around the median spectral index, $\sigma=0.16$. The blue line indicates the integrated spectrum of the halo.} 
\label{hist_halo}
\end{figure}  

%############################################################################################################   
\subsection{X-ray emission}
\label{Xraymorphology}
%############################################################################################################   

X-ray observations of Abell 2256 suggest a complex dynamical state. The cluster has two bright peaks in the X-ray surface brightness distribution corresponding to a primary cluster (main component) and a subcomponent (``mouth'') to the west, see Fig.\,\ref{halo_Xray} for labeling. The main component has a prominent peak (P1). The peak in the main cluster is actually the center of the cluster (core), while the other peak is about 250\,kpc to the west. Optical observations also suggest a third poorer subcomponent to the north of the cluster which is infalling onto the main component from the northeast \citep{Sun2002,Miller2003}. 

In Fig.\,\ref{halo_Xray}, we compare the X-ray morphology of the cluster with that of radio emission at 675\,MHz (top panel) and 144\,MHz (bottom panel). The radio halo emission at 675 and 144\,MHz largely follows the X-ray surface brightness distribution. In both radio and X-ray, the main component is elongated along the southeast-northwest direction.  Radio emission from the halo extends over the entire region of the detected X-ray emission. 

From Figs.\,\ref{halo_Xray} and \ref{unsharp}, it is evident that the innermost bright halo core (P1) coincides with the bright X-ray emission in the main component. Within P1, there are two X-ray peaks but none of them coincides with the halo peak in the radio. This suggests that the main component core does not coincides with the peak in the radio halo core. Radio emission from the subcomponent is dominated by three bright radio sources (A, B, and C). Deep X-ray studies provide evidence for five discontinuities in X-ray surface brightness: three cold fronts (CF1, CF2, and CF3) and two shock fronts \citep{Ge2020,Breuer2020}. These discontinuities (except the north-west shock front) are labeled in Fig.\,\ref{halo_Xray}.

%%%%%%%%%%%%%%%%%%%%%%%%%%%
%Figure 8
%%%%%%%%%%%%%%%%%%%%%%%%%%%

\begin{figure*}[!thbp]
    \centering
      \includegraphics[width=1.0\textwidth]{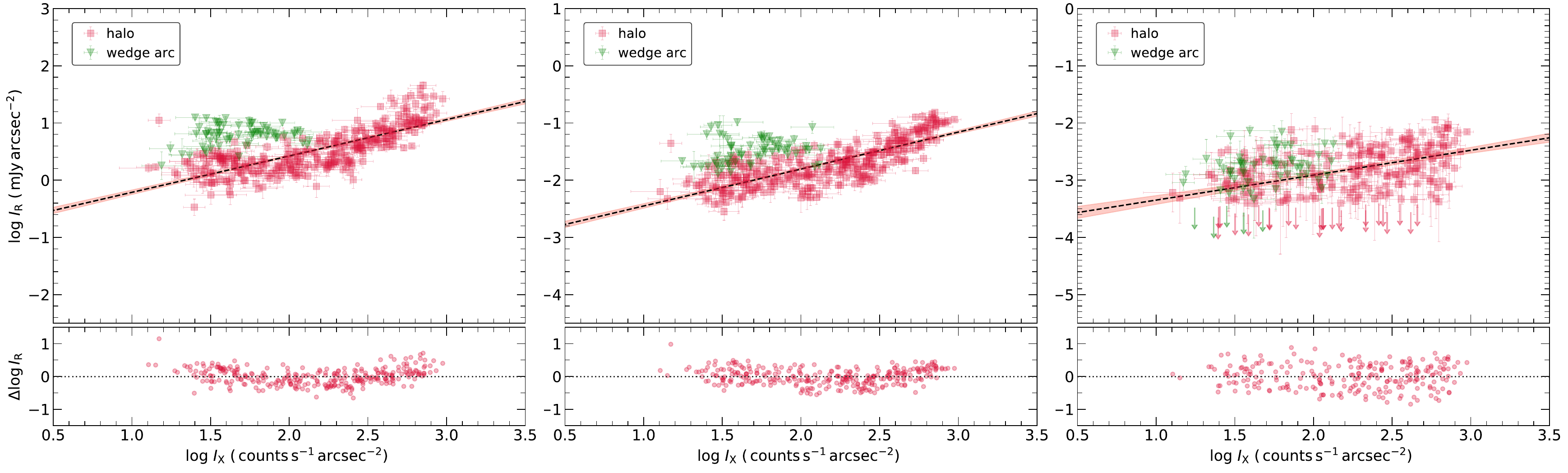}    
 \caption{Radio versus X-ray surface brightness correlation for the halo in Abell 2256 at 144\,MHz (Left), 675\,MHz (Middle), and 1.5 GHz (Right). The surface brightness is extracted from square-shaped boxes (width of 23\,kpc) and in a ``common region'' visible across all three frequencies. The X-ray surface brightness is extracted from the \textit{XMM-Newton} (0$.5-2$\,keV band) image smoothed with a Gaussian FWHM of $6\arcsec$ to have a good signal to noise ratio in the fainter outer regions. The radio surface brightness is extracted from radio maps at 20\arcsec resolution. The {\tt Linmix} best-fit relations for the halo data points are depicted as black dashed lines. Squared-shape boxes represent cells where both the radio and X-ray surface brightness is greater than $3\sigma$.  The $2\sigma$ upper limits are indicated by arrows. The lower panel shows the residuals of $\log I_{\rm R}$ and $\log I_{\rm X}$ with respect to the {\tt Linmix} best fit line for the halo region only. Green data points are extracted from the wedge arc.}
      \label{IRX}
\end{figure*}

In the top-right panel of Fig\,\ref{unsharp}, we show an unsharp-masked \textit{Chandra} image of Abell 2256 obtained with Gaussian smoothing of $\sigma_1=4\arcsec$ and $\sigma_2=40\arcsec$. Unsharp-masked maps provide a way to investigate the presence of (sharp) discontinuities in the X-ray surface brightness images. The image highlights the presence of three cold fronts (CF1, CF2, and CF3) reported in the inner region of the halo. The positions of these cold fronts are indicated by magenta-colored arcs. It is clear that the bright radio peak P1, in the main component, is confined by the innermost cold front CF3. The morphological similarities between the halo core in radio and X-ray are evident. Source D is an FRI radio source \citep{Fanaroff1974}, contained within the isophotes of the optical galaxy \citep{Owen2014}. Low-frequency images do not show any morphological connection between source D and the halo core. Within the inner region of the halo, there are three bright cluster galaxies, namely G1, G2, and G3, see bottom-right panel of Fig.\,\ref{unsharp}. G1 and G3 are located in the main component, but we do not find any radio counterpart in our deep observations but they could have been active in the past and therefore may be providing fossil plasma. Moreover, the peak in the radio halo emission is not coincident with G1 or G3.  G2 belongs to the secondary component \citep{Ge2020}.

Recently, \cite{Breuer2020} identified a step-like low-frequency radio surface brightness feature, detected at 325 MHz, that coincides with the long-tail-like X-ray feature (i.e., shoulder cold front see Fig.\,\ref{unsharp}). Compared to \cite{Breuer2020}, our low-frequency radio observations are highly sensitive (by a factor of four at about 325 MHz), but we do not find any clear radio feature that coincides with the shoulder cold front. The radio emission indeed shows a `step' like morphology but is not really tracing the long X-ray tail. 

The emission from the halo extends further to the south, where the X-ray emission is fainter. Recently, \cite{Ge2020} detected a shock front of Mach number $\mathcal{M_{\rm T}}=1.54\pm0.05$ through a temperature jump at the southern edge of the halo (shown in red in Fig.\,\ref{halo_Xray}). As mentioned above, in radio maps the southern edge of the halo follows a V-shaped morphology. This V-shaped edge in the radio is spatially coincident with the southern shock front in the X-ray image. A similar type of morphology is observed in the case of the Toothbrush cluster \citep{vanWeeren2016a}, Bullet cluster \citep{Shimwell2014}. Abell 754 \citep{Macario2011}, Coma cluster \citep{Bonafede2022},  Abell 520 \citep{Markevitch2005,Vacca2014}.   

%############################################################################################################   
\section{Radio halo analysis}
\label{halo_anaylsis}
%############################################################################################################   

\subsection{Integrated spectrum }
\label{halo_int}
To obtain the integrated radio spectrum of the halo, we measure flux densities from the $20\arcsec$ resolution radio images created with a common inner \textit{uv}-cut of $0.1\rm \,k\lambda$ (here, $0.1\rm \,k\lambda$ is the minimum well-sampled uvdistance in the uGMRT Band 4 data),  uv-tapering, and ${\tt robust} =-0.5$. The region used for measuring the halo flux densities is shown in Fig.\,\ref{halo_index} (top right). The flux densities of unrelated compact sources were  measured manually and subtracted from the total halo flux density. Additionally, the entire region covering sources A, B, and C was excluded. Also, the  arc-shaped wedge structure to the east is initially considered a separate structure (discussed in Sect.\,\ref{radio-versus-xray}). 

%%%%%%%%%%%%%%%%%%%%%%%%%%%%%%%%%%%%%%%%%%%%%%%%%%%%%%%%%%%%%%%%%
% Table - spectral properties of the diffuse emission sources 
%%%%%%%%%%%%%%%%%%%%%%%%%%%%%%%%%%%%%%%%%%%%%%%%%%%%%%%%%%%%%%%%%
\setlength{\tabcolsep}{10pt}
\begin{table*}[!htbp]
\caption{Properties of the halo, wedge arc and source AI.}
\centering
\begin{threeparttable} 
\begin{tabular}{*{7}{c}}
\hline \hline
\multirow{1}{*}{Source} &\multirow{1}{*}{LOFAR (144\,MHz)} & \multicolumn{2}{c}{uGMRT (300-850\,MHz)} & \multicolumn{1}{c}{VLA(1-4\,GHz)}&LLS& \multirow{1}{*}{$\alpha$}$^{\dagger\dagger}$\\
 \cline{3-5} 

& $S_{\rm144\,MHz}$ &$S_{\rm350\,MHz}$&$S_{\rm675\,MHz}$&$S_{\rm1.5\,GHz}$&&\\
  & (mJy) & (mJy) & (mJy)&  (mJy) &(Mpc)\\
  \cline{2-3} \cline{4-5}\cline{6-7}
  \hline 
% flux after removing continuation from point sources   
halo &$912\pm120$ &$217\pm26$&$75\pm8$&$20\pm2$& $\sim0.77$&$-1.63\pm0.03$\\ 
arc &$150\pm20$ &$30\pm7$&$15\pm2$&$3.0\pm0.5$& $-$&$-1.62\pm0.04$\\ 
halo+wedge arc &$1062\pm140$ &$247\pm33$&$90\pm10$&$23\pm3$& $-$&$-1.63\pm0.03$\\ 
AI &$41\pm5$ &$15\pm2$&$4.3\pm0.2$&$-$& $\sim0.46$&$-$\\ 
\hline 
\end{tabular}
\begin{tablenotes}[flushleft]
\footnotesize
\item{\textbf{Notes.}} Flux densities were extracted from $20\arcsec$ resolution radio maps created with ${\tt roboust}=-0.5$ and an inner \textit{uv}-cut of $0.1\rm k\lambda$. Absolute flux density scale uncertainties are assumed to be 10\% for LOFAR and Band3, 5\% for uGMRT Band4, and 2.5\% for VLA L-band data. $^{\dagger}$The LLS measured at 675\,MHz; $^{\dagger\dagger}$ the integrated spectral index obtained by fitting a single power law fit. 
\end{tablenotes}
\end{threeparttable} 
\label{Tabel:Tabel2}   
\end{table*} 
  
The resulting integrated spectrum of the halo is shown in the left panel of Fig.\,\ref{halo_index}. The overall spectrum of the halo can be described by a power law. The integrated emission from the halo between 144\,MHz and 1.5\,GHz has an ultra steep spectral index of $-1.63\pm0.03$. The halo has a much steeper spectrum than the large relic which shows a spectral index of  $-1.07\pm0.02$ (paper I).  The steep spectral index of the halo is in agreement with previous studies \citep{Brentjens2008,vanWeeren2012b}. There are only a handful of radio halos with such high-sensitivity radio observations over a wide frequency range, namely the halos in MACS\,J0717.5$+$3745 \citep{Perley2013,Rajpurohit2021b}, 1RXS\,J0603.3$+$4214  \citep{vanWeeren2016a,Rajpurohit2018,Rajpurohit2020a}, CIZA\,J2242.8$+$5301 \citep{vanWeeren2010,Hoang2017,Gennaro2018}, Coma \citep{Bonafede2022}, Abell S1063 \citep{Xie2020}, Abell 2744 \citep{Rajpurohit2021c}, and MACS\,J1149.5$+$2223 \citep{Luca2021}. Out of these eight (including Abell 2256) radio halos mentioned, most show a spectral index steeper than about $-1.4$. Three of these halos (Coma, Abell S1063, and MACS\,J0717.5$+$3745) show a high frequency spectral steepening, while the rest follow a single power-law spectrum. Detailed studies of some radio halos with power law spectra have revealed a complex distribution; subregions exhibit different spectral behavior and scattering in the spectral index distribution \citep{Rajpurohit2021b,Rajpurohit2021c}.

Turbulent reacceleration models predict that a significant fraction of the population of radio halos have ultra steep spectrum \citep[e.g.,][]{Brunetti2008}. In these models, provided that homogeneous conditions apply in the emitting volume, we expect a spectral steepening at higher frequencies \citep{Cassano2006}. The fact that we do not detect a clear steepening in the integrated spectrum of the Abell 2256 halo may imply that the emitting volume is not homogeneous. Perhaps the subregions of the halo show different spectral behavior, but the overall spectrum results in the power-law behavior, as recently reported for the halo in Abell 2744 \citep{Rajpurohit2021c}.

To investigate this, we divide the halo into four subregions, core, region1, region2, and region3. These regions are shown in the bottom right panel of Fig.\,\ref{halo_index}: the core is the innermost region and region3 the outermost. The regions that are masked are shown in magenta. The resulting spectra are  shown in Fig.\,\ref{halo_index} (bottom left panel). The subregions of the halo show different spectral slopes. The core can be well described by a power law spectrum with slope $-1.60$. The spectrum of region2 can also be described by a power law with slope $-1.44$. The core region is  steeper compared to region2. There is a clear high frequency steepening in region1 and region3.  This implies that a general power-law emission spectrum, observed for many halos, does not necessarily imply that electrons have everywhere in the halo the same power-law distribution  \citep{Rajpurohit2021c}. 

Excluding the core region, it is clear that there is a radial steepening in the halo moving to larger radii. The synchrotron frequency (or steepening) in the turbulent reacceleration model \citep{Cassano2005} is given by:
\begin{equation}
\nu_s \propto \frac{\tau_{\rm acc}^{-2} B}{(B^2 + B_{\rm cmb}^2)^2},    
\end{equation}
where $B$ is the magnetic field and $\tau_{acc}$ is the minimum acceleration time in the emitting volume. Therefore, a radial steepening is expected from the decline of the magnetic field (for constant $\rm \tau_{acc}$), at least at large distances where presumably $B < B_{\rm cmb}/\sqrt{3}$ or due to a decline of $\tau_{\rm acc}^{-2}$ (i.e., less acceleration at larger distances). However, this assumes that local conditions change only as a function of radius.

\setlength{\tabcolsep}{8pt}
\begin{table*}
\caption{{\tt Linmix} fitting slopes and Spearman ($r_{s}$) and Pearson ($r_{p}$) correlation coefficients of the data for Figs.\,\ref{IRX} and \ref{IRX_main}.}
\centering
\begin{threeparttable} 
\begin{tabular}{ c  c  c  c c c  c  c  c  c }% c | c}
 \hline  \hline  
\multirow{1}{*}{} & \multirow{1}{*}{$\nu$} &  \multicolumn{4}{c|}{$3\sigma$} & \multicolumn{4}{c}{$2\sigma$ as upper limits}\\
 \cline{2-10}
&& $b$ &   $\sigma_{\rm int}$ &   $r_{s}$   & $r_{p}$ & $b$ & $\sigma_{\rm int}$ & $r_{s}$ & $r_{p}$\\
  \hline  
Halo&$144\,\rm MHz$& $0.64\pm0.02$& $0.08\pm0.01$& $0.79$& $0.76$& $0.64\pm0.02$& $0.08\pm0.01$& $0.79$& $0.76$\\
(common region)&$675\,\rm MHz$& $0.64\pm0.03$& $0.05\pm0.01$& $0.81$& $0.76$& $0.65\pm0.03$& $0.05\pm0.01$& $0.81$& $0.76$\\ 
&$1.5\,\rm GHz$& $0.37\pm0.05$& $0.07\pm0.01$& $0.55$& $0.39$& $0.44\pm0.05$& $0.07\pm0.05$& $0.54$& $0.39$\\ 
\hline
Halo+wedge arc&$144\,\rm MHz$& $0.43\pm0.03$& $0.09\pm0.01$& $0.56$& $0.54$& $0.43\pm0.03$& $0.09\pm0.01$& $0.56$& $0.54$\\
(common region)&$675\,\rm MHz$& $0.44\pm0.03$& $0.08\pm0.01$& $0.60$& $0.56$& $0.44\pm0.03$& $0.08\pm0.01$& $0.60$& $0.56$\\ 
&$1.5\,\rm GHz$& $0.28\pm0.04$& $0.07\pm0.01$& $0.46$& $0.33$& $0.33\pm0.04$& $0.07\pm0.01$& $0.46$& $0.27$\\ 
\hline
&144\,MHz& $0.60\pm0.01$& $0.06\pm0.01$& $0.89$& $0.86$& $0.70\pm0.02$ &$0.15\pm0.01$ &0.87&0.85 \\ 
Halo&675\,MHz & $0.60\pm0.02$& $0.05\pm0.01$& $0.87$& $0.80$&$0.72\pm0.02$&$0.08\pm0.01$&0.88&0.84 \\ 
&$1.5\,\rm GHz$& $0.37\pm0.05$& $0.07\pm0.01$& $0.55$& $0.39$& $0.44\pm0.05$& $0.07\pm0.05$& $0.54$& $0.39$\\ 
  \hline  
&$144\,\rm MHz$&  $0.61\pm0.02$& $0.09\pm0.01$& $0.83$& $0.80$& $0.73\pm0.02$& $0.15\pm0.01$& $0.82$& $0.80$\\
Halo+wedge arc&$675\,\rm MHz$& $0.57\pm0.02$& $0.08\pm0.01$& $0.80$& $0.70$& $0.73\pm0.02$& $0.12\pm0.01$& $0.83$& $0.79$\\ 
&$1.5\,\rm GHz$& $0.28\pm0.04$& $0.07\pm0.01$& $0.46$& $0.33$& $0.33\pm0.04$& $0.07\pm0.01$& $0.46$& $0.27$\\ 
\hline 
\end{tabular}
\end{threeparttable} 
\label{fit}   
\end{table*}

%############################################################################################################   
\subsection{Radio power versus mass relation}
%############################################################################################################   

Using the flux densities measured at 1.5\,GHz and 144\,MHz, we estimate the total radio power of the halo. We note that the arc-shaped wedge is also considered part of the halo. The total rest-frame radio powers of the halo are $P_{1.4\,\rm GHz,\,lower\,limit}=(1.9\pm0.2)\times10^{23} \,\rm W\,Hz^{-1}$ and $P_{150\,\rm MHz,\,lower\,limit}=(8.9\pm0.8)\times10^{24} \,\rm W\,Hz^{-1}$. Our new measurement of the halo yields a 1.4\,GHz radio power that is at least 4 times lower than previously reported from \cite{Clarke2006}, namely $8.2\times10^{23} \,\rm W\,Hz^{-1}$. We note that \cite{Clarke2006} used a larger and different area for the measurement of the flux density of the halo plus the uncertainty in the flux density was on the order of 20\%. Moreover, their flux density value may be affected by unresolved unrelated sources.

%%%%%%%%%%%%%%%%%%%%%%%%%%%
%Figure 9
%%%%%%%%%%%%%%%%%%%%%%%%%%%

 \begin{figure*}[!thbp]
    \centering
          \includegraphics[width=0.48\textwidth]{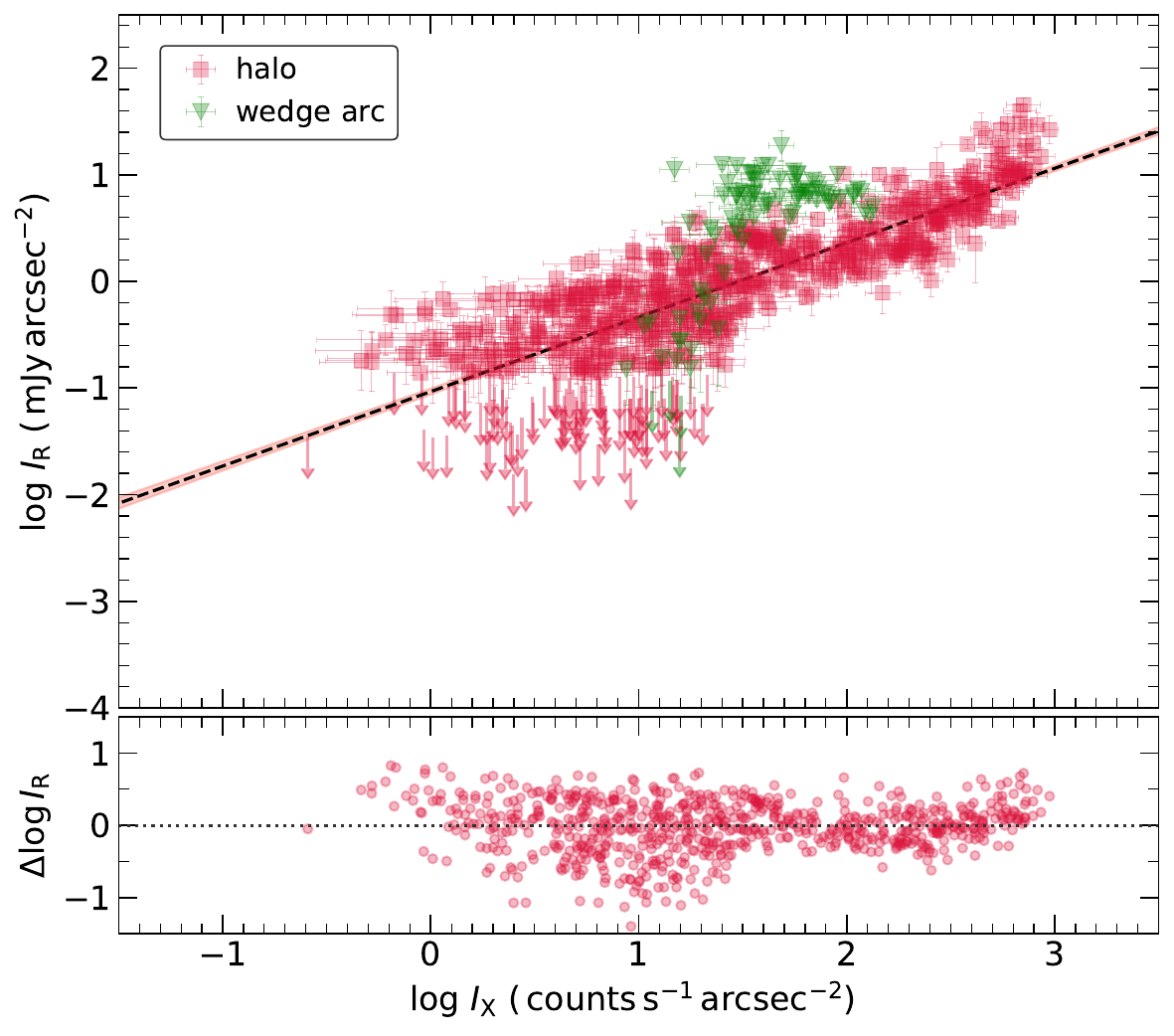}   
      \includegraphics[width=0.48\textwidth]{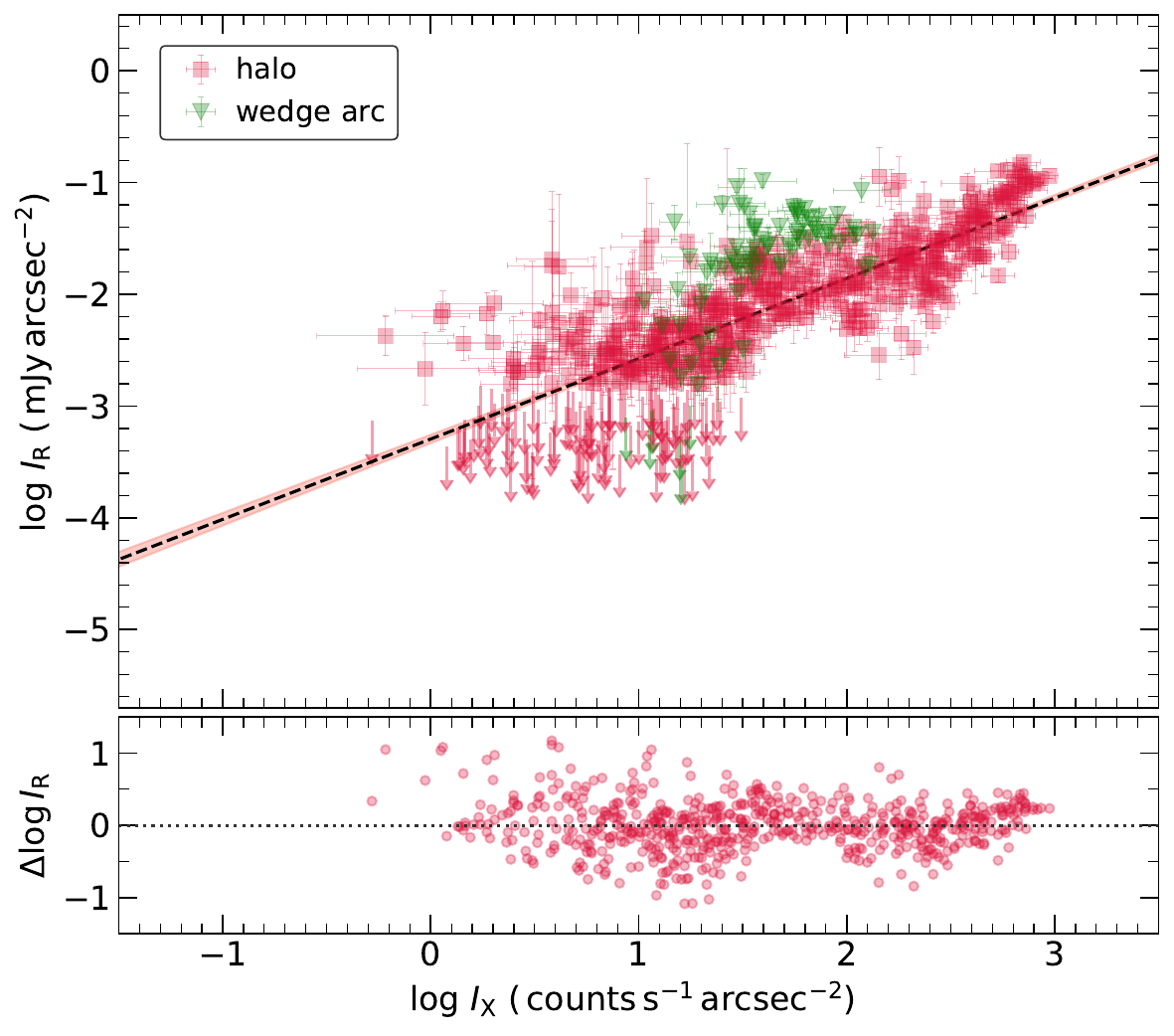}    
 \caption{Radio versus X-ray surface brightness correlation for the halo in Abell 2256 at 144\,MHz (left) and 675\,MHz (Right). Images and fitting is the same as in Fig.\,\ref{IRX} but including all regions where both radio and X-ray surface brightness exceeds $2\sigma$ (emission below $3\sigma$ is shown as upper limits). Correlation between the radio and X-ray surface brightness across the halo region gets tighter when including all the halo emission at low frequencies.  The best-fitted correlation slopes are $b_{\rm 144\,MHz}=0.87\pm0.02$,  $b_{\rm 675\,MHz}=0.88\pm0.03$ in the halo region. The data points from the wedge arc are not scattered but are above the halo correlation.}
      \label{IRX_main}
\end{figure*}  

%%%%%%%%%%%%%%%%%%%%%%%%%%%
%Figure 10
%%%%%%%%%%%%%%%%%%%%%%%%%%%

 \begin{figure}[!thbp]
    \centering
          \includegraphics[width=0.49\textwidth]{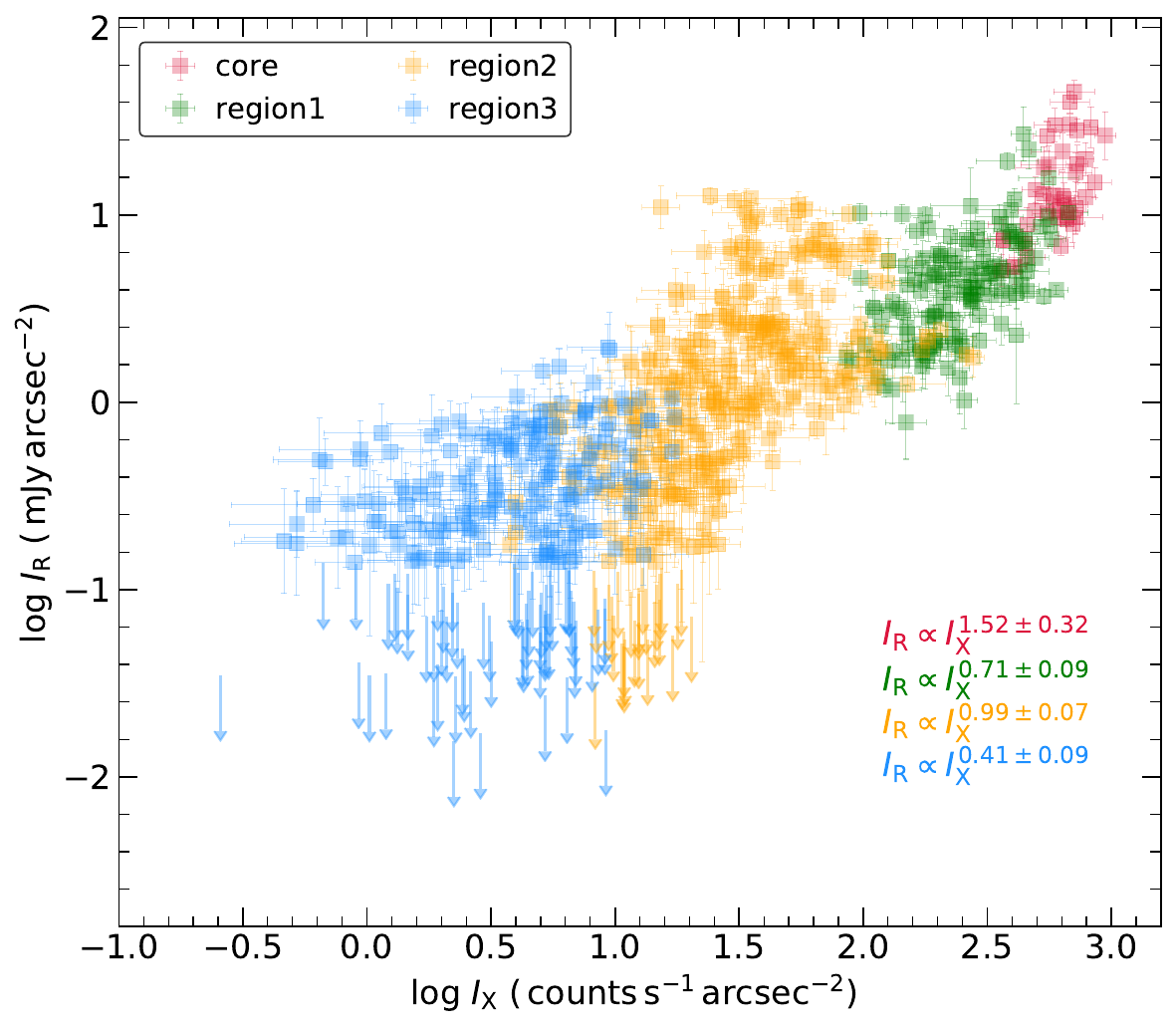}      
 \caption{The radio versus X-ray surface brightness distribution at 144\,MHz (similar to Fig.\,\ref{IRX_main} left panel) in the inner and outer regions of the halo; color coded with different circles (see Fig.\,\ref{halo_index} bottom right panel). Blue and green circles (region3) covers the outermost region while the core that of the innermost region of the halo. The subregions of the halo show different correlation slopes.}
      \label{subregions}
\end{figure} 

The radio halo is very likely seen in projection with sources A, B, and C, so the above estimated radio power provides a lower limit. Moreover, part of the halo emission could be seen in projection with the relic. However, the large-scale structure of Abell 2256 makes it very difficult to determine the exact northern boundary of the halo. As discussed in Sec.\,\ref{Xraymorphology}, the radio halo morphology nicely traces the X-ray morphology. Therefore, we take the northwest boundary of the halo using the X-ray distribution at that location. To estimate the upper limit of the halo power, we extrapolated the halo flux density in the area covering sources A, B, C, and the northwest part of the relic by using the average flux from the halo per unit surface area outside that region. In this way, the total halo flux density values at 144 MHz and 1.5 GHz are 1570 mJy and 36 mJy, respectively. This gives the radio power of $P_{1.4\,\rm GHz,\,upper\, bound}=3.1\times10^{23} \,\rm W\,Hz^{-1}$ and $P_{150\,\rm MHz,\,upper\, bound}=1.3\times10^{25} \,\rm W\,Hz^{-1}$. We also checked the halo flux density using the {\tt halo-FDCA} package \citep{Boxelaar2021} and the resulting values are comparable (Osinga et al. in prep.). 

It is well known that the radio power (at 1.4\,GHz and 150\,MHz) of halos show a well defined correlation with the mass and X-ray luminosity of the host cluster \citep{Cassano2013,vanWeeren2020,Duchesne2021,Cuciti2021}.  With our new radio power (considering upper bound), the halo in Abell 2256 is a factor of 4 below this correlation at 1.4\,GHz \citep{Cuciti2021}. The halo falls into the category of underluminous halos in the known 1.4\,GHz radio power versus mass relation \citep{Cassano2013}.  To the best of our knowledge, underluminous radio halos are also reported in Abell 1451 \citep{Cuciti2018}, ZwCl\,0634$+$47 \citep{Cuciti2018}, Abell 1430 \citep{Hoeft2020}, Abell 3667 \citep{deGasperin2022}, Abell 1689 \citep{Vacca2011}, Abell 2218 \citep{Giovannini2000} and Abell 3266 \citep{Riseley2022b}.  We emphasize that, compared to these known underluminous halos, the one in Abell 2256 has a reliable characterization of its spectrum. Moreover, the radio power of halos at 1.4\,GHz scales with the X-ray luminosity of the hosting clusters. Ultra-steep spectrum radio halos are expected to be underluminous with respect to this correlation \citep{Cassano2010a}. Moreover, MHD simulations suggest that during the formation and evolution of the halos, the halo can be underluminous or luminous  \citep{Donnert2013}: steep spectrum radio halos in the very early or very late stages of their lifetimes are expected to be underluminous. This scenario seems plausible for the radio halo in Abell 2256 due to its ultra steep spectrum and low radio power at 1.4\,GHz. Similar to the radio power vs mass relation, the Abell 2256 halo is underluminous also in the radio power vs. X-ray luminosity relation. In contrast to the radio power versus mass relation at 1.4\,GHz, the halo fits well in the radio power versus mass relation at 150\,MHz \citep{vanWeeren2020,Duchesne2021}. This hints that these correlations are frequency-dependent.

%############################################################################################################   
\subsection{Spectral index distribution}
%############################################################################################################   
To check the spectral variations in the halo, we construct a spectral index map of the halo between 144\,MHz and 675\,MHz using the $20\arcsec$ resolution images described in Sect.\,\ref{halo_int}. Since the total extent of the halo decreases at high frequencies, which in turn significantly reduces the area over which we can study the spectral index distribution, we do not use 1.5\,GHz data. Pixels with flux densities below $3\sigma_{\rm rms}$ at both frequencies were blanked.

In the top panel of Fig.\,\ref{halo_spectral_index_map}, we show the spectral index map of the halo emission between between 144\,MHz and 675\,MHz. Excluding unrelated sources, the spectral index, across the halo, varies between $-1.3$ and $-1.8$.  The core is evidently steep ($-1.65$) compared to the surrounding region. This value is consistent with the overall spectrum of the core region. It is plausible that the bright steep core is associated with the previous AGN activity. Thus its spectrum is affected by the external halo emission seen in projection. This would imply that the core spectrum is even steeper than observed  ($-1.60$), since true region 1 is projected on the core. We note that in case the halo core is related to an old AGN lobe, we expect a curved spectrum. In contrast, the halo core follows a power-law spectrum between 144 MHz and 1.5 GHz.

%%%%%%%%%%%%%%%%%%%%%%%%%%%
%Figure 11
%%%%%%%%%%%%%%%%%%%%%%%%%%%

\begin{figure}[!thbp]
\centering
 \includegraphics[width=0.49\textwidth]{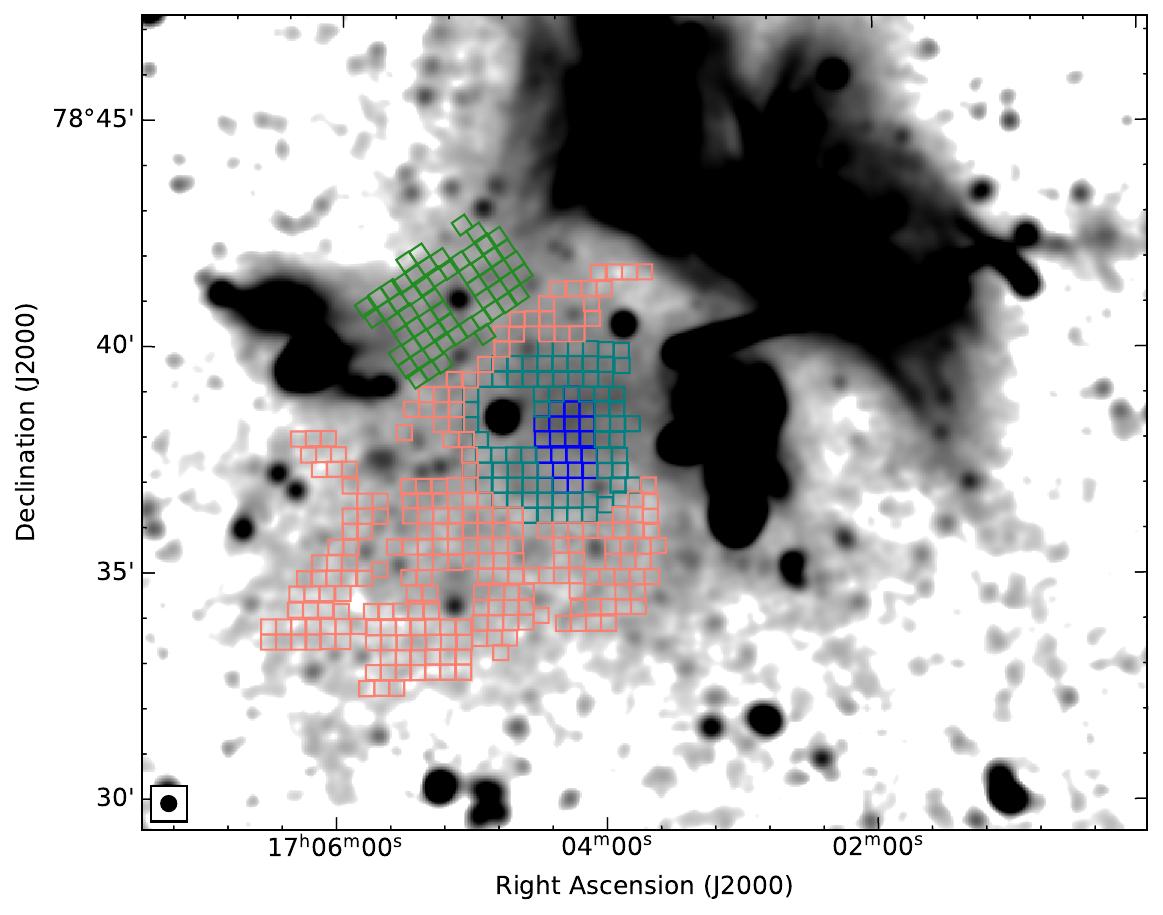}   
 \includegraphics[width=0.49\textwidth]{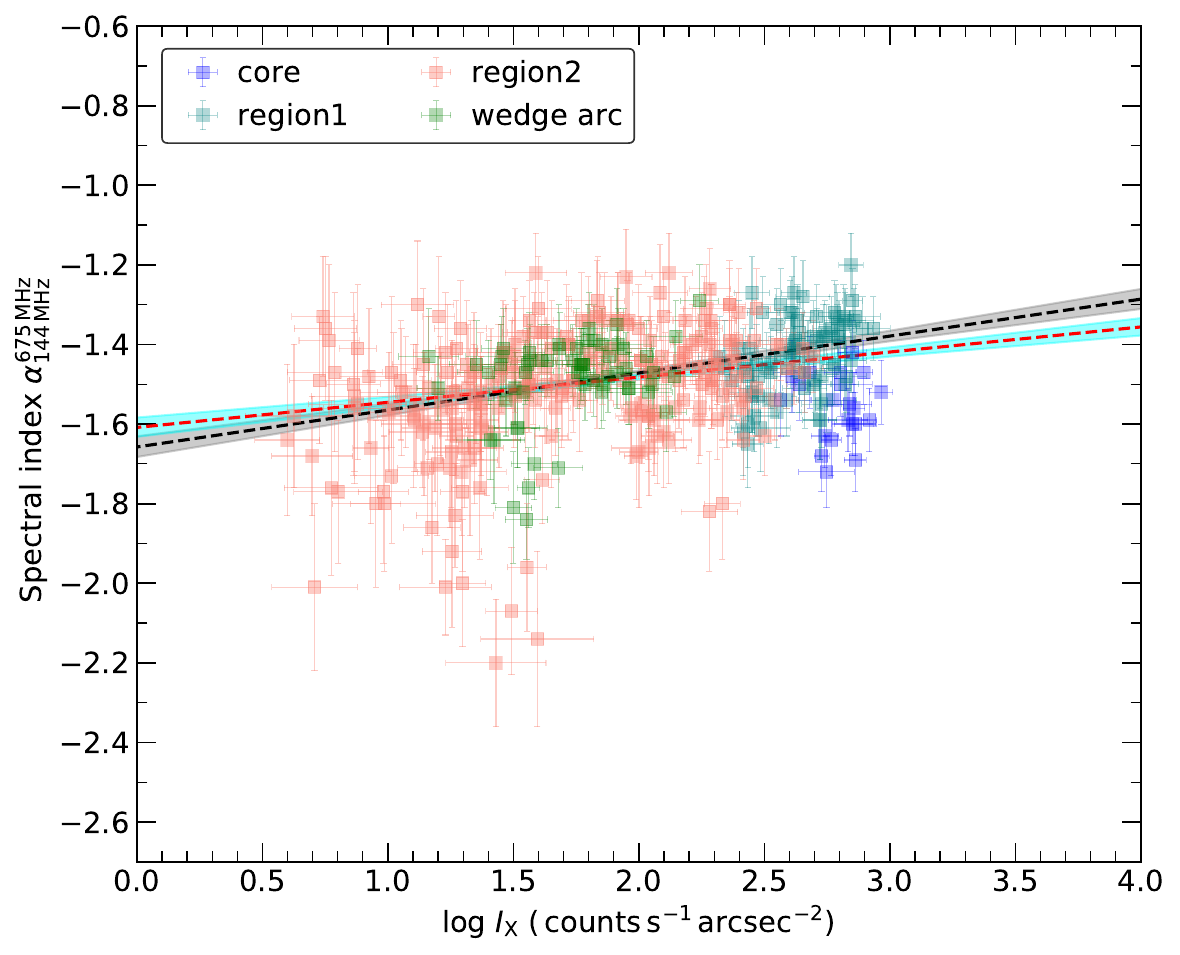}   
 \caption{\textit{Top}: Regions used to obtain the $\alpha-I_{\rm X}$ correlation. The radio spectral index values were extracted between 144\,MHz and 675 MHz. Each box has a width of 20\arcsec corresponding to a physical size of about 23\,kpc. \textit{Bottom}: The spectral index versus X-ray surface brightness relation across the halo in Abell 2256. There is a strong anticorrelation between these two quantities (the black dashed line shows the fitting excluding the core region while the red dashed line includes the core region).}
\label{IX_index}
\end{figure}

In Fig.\,\ref{hist_halo}, we show a histogram of the spectral index in the halo. The distribution is asymmetric, showing a median value of $\langle\alpha\rangle=-1.48$ and a standard deviation of $\sigma=0.16$. If the variations in the spectral index are the result of measurement errors, we expect the median error value to be comparable to the standard deviation. For the halo, we find a median error of $0.08$. This value is about a factor of 2 smaller than the standard deviation, which implies that there are intrinsic small-scale fluctuations in the spectral index across the halo. Fluctuations in the spectral index are also reported for the halos in Abell 2255 \citep{Botteon2020a}, Abell 520 \citep{Hoang2019}, MACS\,J0717.5+3745 \citep{Rajpurohit2021b}, and Abell 2744 \citep{Rajpurohit2021c}.  We emphasize that the fluctuations in the spectral index of the Abell 2256 halo again highlights the contradiction with the overall spectrum, which is a power law. The spatially resolved spectral index map shows a mean spectral index of $-1.50$, which is slightly flatter than the one obtained from the integrated spectrum. However, this is expected since the halo is more extended at 144\,MHz and in those regions the radio surface brightness is below $3\sigma_{\rm rms}$ at 675\,MHz, thus not included in Fig.\,\ref{halo_index}.  Moreover,  the integrated spectrum is brightness weighted, while in the spectral index map each pixel has the same relevance, regardless of its brightness (i.e., signal to noise ratio).

%############################################################################################################   
\subsection{Radio versus X-ray surface brightness}
\label{radio-versus-xray}
%############################################################################################################   

Radio halos often show a point-to-point correlation between radio and X-ray surface brightness \citep{Govoni2001a,Govoni2001b,Shimwell2014,Rajpurohit2018,Hoang2019,Cova2019,Xie2020,Botteon2020,Rajpurohit2021b,Rajpurohit2021c,Luca2021,Duchesne2021,Hoang2021,Bonafede2022,Riseley2022b}. The correlation slope is mostly found to be sublinear in radio halos, i.e., the nonthermal radio emission declines less rapidly than the thermal X-ay emission. Relatively few radio halos have high-quality multi-frequency radio data available. Of these halos, a few show that the correlation slope changes as a function of frequency \citep{Rajpurohit2021b,Hoang2021}. However, there is also a case where the correlation slope remains constant as a function of frequency, the halo in Abell 2744 \citep{Rajpurohit2021c}.

We checked whether there is any correlation between the X-ray and radio surface brightness for the halo in Abell 2256. We used $20\arcsec$ resolution radio maps at 144\,MHz, 675\,MHz and 1.5\,GHz. For X-ray, we use a \textit{XMM-Newton} $0.5-2$ keV image smoothed with a Gaussian of $6\arcsec$ FWHM. First, we create a grid (with square boxes of width $20\arcsec$) that covers the halo region visible at all three frequencies, including the arc-shaped structure on the eastern side of the halo. We exclude areas with discrete radio and X-ray point sources. We include regions where the radio and X-ray surface brightness is above the $2\sigma$ level. The emission below $3\sigma$ is included as upper limits.

\setlength{\tabcolsep}{10pt}
\begin{table}
\caption{{\tt Linmix} correlation slope within the halo subregions of Fig.\,\ref{subregions}.}
\centering
\begin{threeparttable} 
\begin{tabular}{ c c c c}
 \hline  \hline  
& slope (b)& correlation coefficient $(r_{s})$ \\ 
\hline  
core&$1.51\pm0.32$&0.64\\
region1&$0.71\pm0.09$&0.56\\
region2&$0.99\pm0.07$&0.61\\
region3&$0.41\pm0.09$&0.18\\
\hline 
\end{tabular}
\end{threeparttable} 
\label{fit_subregions}   
\end{table}

The resulting plots at 144\,MHz, 675\,MHz, and 1.5\,GHz are shown in Fig.\,\ref{IRX}.  Despite the complex distribution of the thermal and nonthermal emission components at the center, there seems to be a positive correlation between the radio and X-ray surface brightness for all three frequencies: higher radio brightness is associated with higher X-ray brightness. To check the strength of the correlation, we fit the data with a power law of the form
\begin{equation}
I_{\rm R}\propto I_{\rm X}^{b},
\end{equation}
where $b$ is the correlation slope. We use the {\tt Linmix} package  \citep{Kelly2007} for fitting.
The Spearmann correlation coefficients at the observed frequencies are $r_{\rm s,\,144 MHz}=0.56$, $r_{\rm s,\,675\,MHz}=0.60$, and $r_{\rm s,\,1.5\,GHz}=0.46$. There is a moderate correlation between the two quantities, but it is not significant at 1.5\,GHz.  The correlation slope is sublinear and more or less constant at 144\,MHz and 675\,MHz, namely $b_{\rm 144\,MHz}=0.43\pm0.03$ and $b_{\rm 675\,MHz}=0.44\pm0.03$.

When excluding the arc-shaped wedge, we find that for the rest of the data points, the radio and X-ray surface brightness is correlated at 144\,MHz and 675\,MHz with correlation coefficient of ${\rm r_{s,\,144\,MHz}}=0.79$ and $r_{\rm r_{s},\,675 MHz}=0.81$, respectively. At high frequencies, the correlation is comparatively weak, namely ${\rm {r_{s,\,1.5\,GHz}}}=0.54$. Similar to other known halos, the correlation slope is sublinear in the Abell 2256 halo, see Table\,\ref{fit}. However, unlike some other halos, the correlation slope is flatter at high frequencies. Similar trends are observed  for the halos in CLG\,0217+70 \citep{Hoang2021} and Abell 520 \citep{Hoang2019}. For the halo in CLG\,0217+70, a flat correlation slope at high frequency is caused by X-ray surface brightness discontinuities at the edges of the halo that indicate regions where re-acceleration take place.

%%%%%%%%%%%%%%%%%%%%%%%%%%%
%Figure 12
%%%%%%%%%%%%%%%%%%%%%%%%%%%

 \begin{figure}[!thbp]
    \centering
          \includegraphics[width=0.48\textwidth]{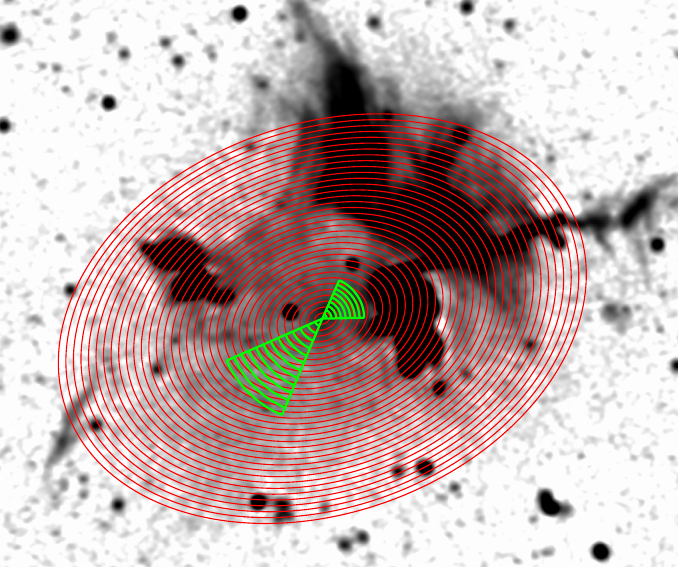}      
 \caption{LOFAR 144 MHz radio image of the Abell 2256 cluster at $20\arcsec$ resolution overlaid with the red ellipse (with a major axis radius of 700\,kpc) that is used to extract radial profiles shown in Fig\,\ref{radial_profiles}. The green sectors show the radial profiles extracted across the  cold front CF2 and CF3. The width of the each annulus is 23\,kpc.}
      \label{profile_regions}
\end{figure}

%%%%%%%%%%%%%%%%%%%%%%%%%%%
%Figure 13
%%%%%%%%%%%%%%%%%%%%%%%%%%%

\begin{figure}[!thbp]
\centering
\includegraphics[width=0.49\textwidth]{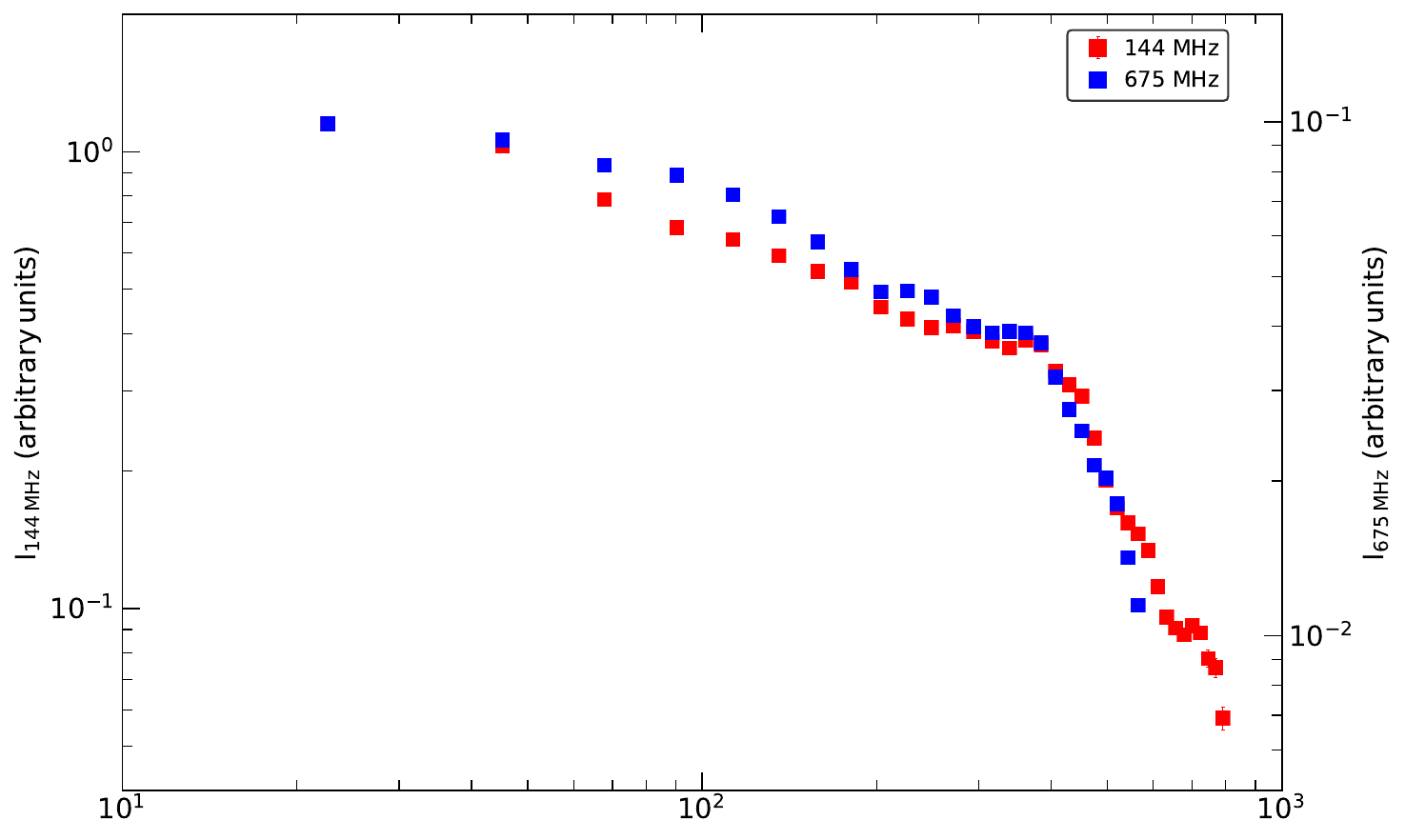} 
\includegraphics[width=0.49\textwidth]{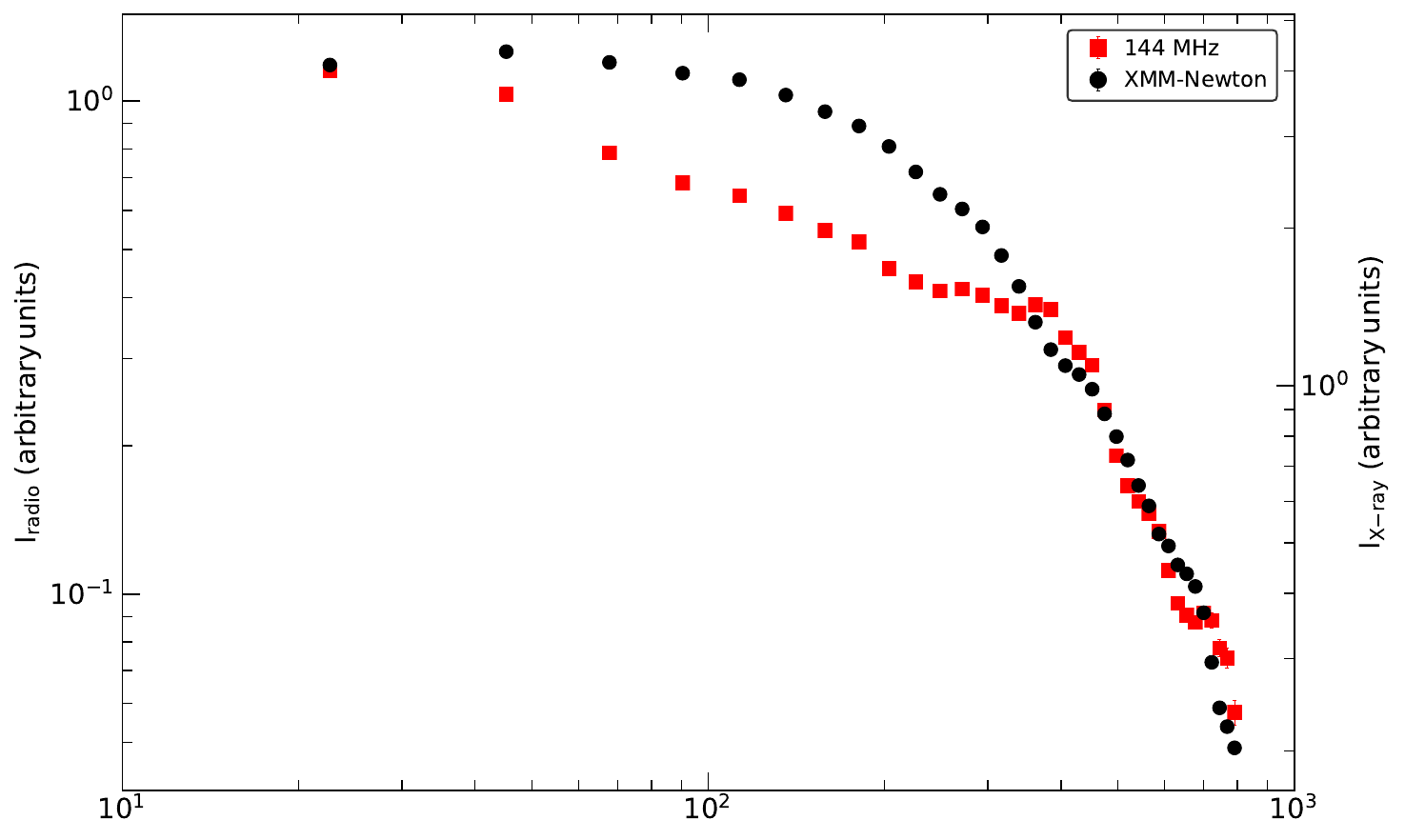}   
\includegraphics[width=0.49\textwidth]{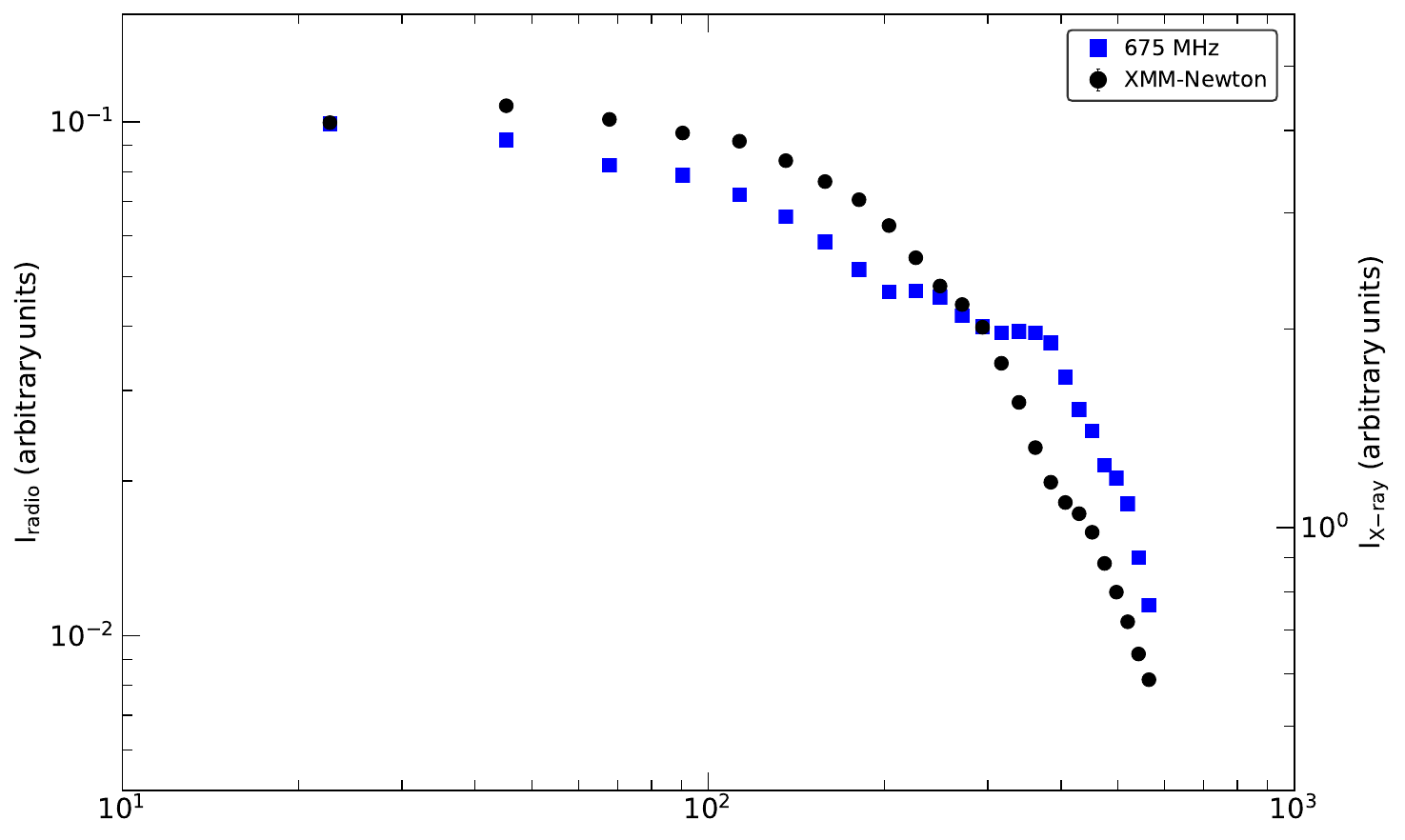}  
\includegraphics[width=0.49\textwidth]{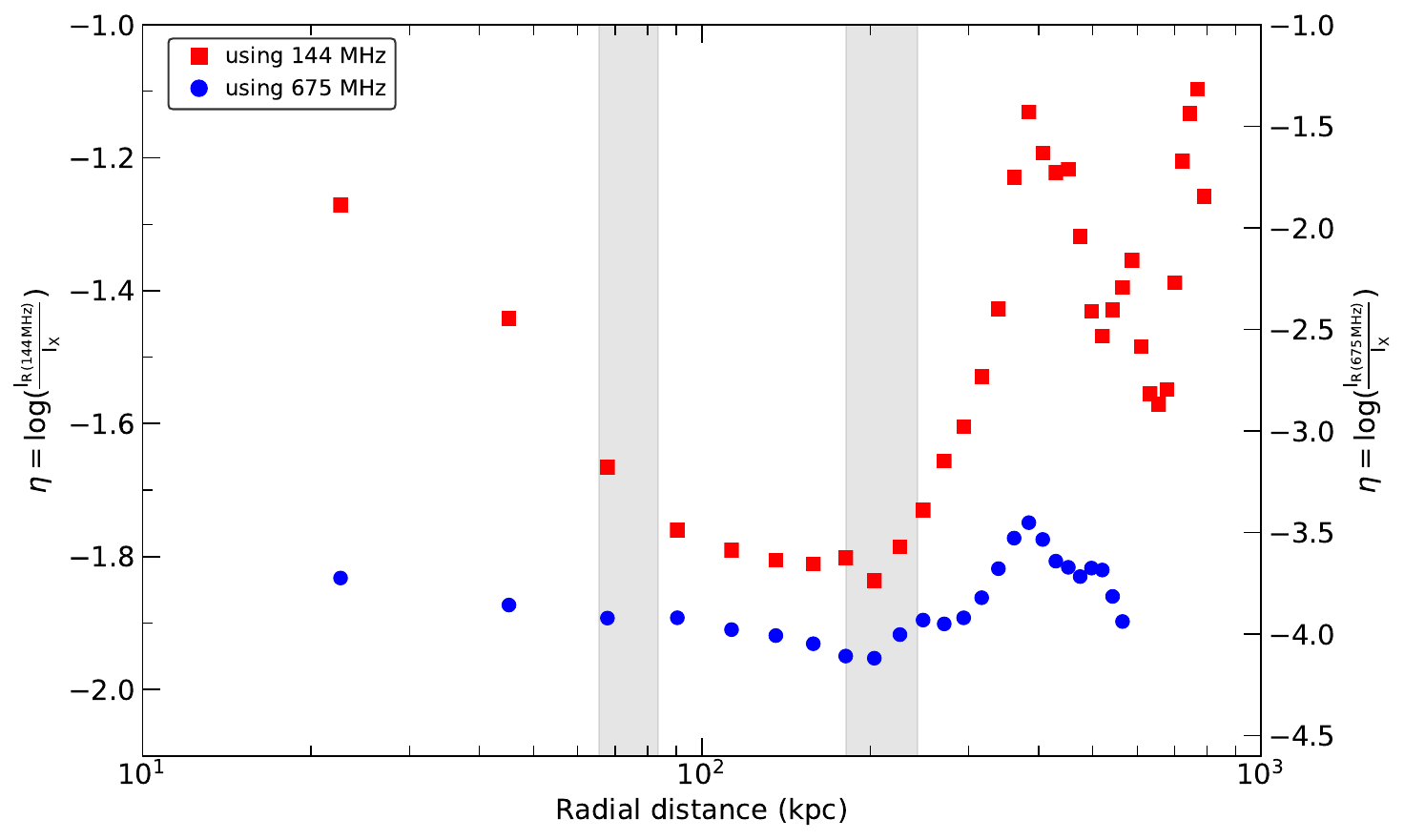}  
\caption{Radial profiles extracted from radio and X-ray maps. The 144 MHz and 675 MHz profiles seem more or less similar (Top panel). However, the radio profiles seem steeper than the X-ray (Middle panels). The bars represent only statistical errors and do not appear because they are smaller than the data points. The $\eta$ profile (Bottom panel) extracted by dividing radio (144\,MHz and 675\,MHz) and X-ray surface brightness. The gray-shaded areas show the location of cold fronts (CF3 and CF2). The region used for extracting the profiles is shown with red in Fig.\,\ref{profile_regions}.}
\label{radial_profiles}
\end{figure}  

Four X-ray surface brightness edges have been reported within the halo in Abell 2256 \citep{Ge2020}. Three of them are cold fronts detected in the inner regions, whereas the fourth is a shock front found to the southeast of the halo. When considering the data points of the wedge arc as part of the halo, the radio versus X-ray correlation is not that strong at all three frequencies; see Table\,\ref{fit}. As shown in Fig.\,\ref{IRX}, the data points from the wedge arc are mostly clustered above the correlation (in particular at 144\,MHz and 675\,MHz) when considering only the region of the halo visible at all frequencies and  the correlation is moderate.  However, when considering the entire halo emission, i.e., including the emission  $>3\sigma$ level ($2\sigma$ as upper limits) at 144\,MHz and 650\,MHz, there is a strong correlation. We emphasize that the data points from the wedge arc lie above the correlation but do not scatter significantly from the halo data points  (see Fig.\,\ref{IRX_main}). Moreover, the correlation slope is more or less the same with and without the wedge arc.

As the total extent of the halo emission is large below 1.5\,GHz, we created another grid that covers the entire halo region ($>3\sigma$) at 144\,MHz and 675\,MHz. The resulting plots are shown in Fig.\,\ref{IRX_main}. In these plots, a large region with less dense X-ray regions can be seen. Moreover, compared to Fig.\,\ref{IRX}, there is a strong correlation between radio and X-ray surface brightness with a correlation coefficient of ${\rm r_{s,\,144\,MHz}}=0.88$ and $r_{\rm s,\,675 MHz}=0.87$. 

Following \cite{Bonafede2022}, we study the correlation of radio and X-ray surface brightness in the inner and outer regions of the halo. Since the halo is more extended toward low frequencies, we use the LOFAR 144\,MHz map to measure the radio surface brightness.  We repeat the analysis described above, i.e.,  extract the radio and X-ray surface brightness within square-shaped boxes of size 23\,kpc but this time in the subregions: core, region1, region2, and region3. The resulting plot is shown in Fig\,\ref{subregions}. We find that the halo slope changes drastically from the core to region3.  Within the core region, the correlation slope is $1.51\pm0.32$ while in the outermost region (region3)  $0.41\pm09$. The slopes for region1 and region2 are $0.71\pm0.09$ and $0.99\pm0.07$, respectively. This indicates that the slope is superlinear in the halo core and sublinear in the outermost fainter region. It worth noticing that the opposite trend has been found for the halo in the Coma cluster \citep{Bonafede2022}, possibly highlighting different physical conditions in these two systems. A superlinear slope is found in the inner region followed by flattening in the outermost region for some cool core clusters that host a hybrid halo, that is, a mini halo and halo-type component, for example in RXC J1720.1+2638 \citep{Biava2021}, MS 1455.0+2232 \citep{Chris2022} and Abell 1413 (Lusetti et al, in prep).  We emphasize that there is no systematic trend in the correlation slope for the halo in Abell 2256 as we move toward  the outer regions. The presence of different slopes in the halo subregions (core, region1, region2, and region3) can be induced by several processes and their interplay, as discussed in Sec.\,\ref{radial_profile_section}.

%%%%%%%%%%%%%%%%%%%%%%%%%%%
%Figure 14
%%%%%%%%%%%%%%%%%%%%%%%%%%%

 \begin{figure}[!thbp]
    \centering
         \includegraphics[width=0.49\textwidth]{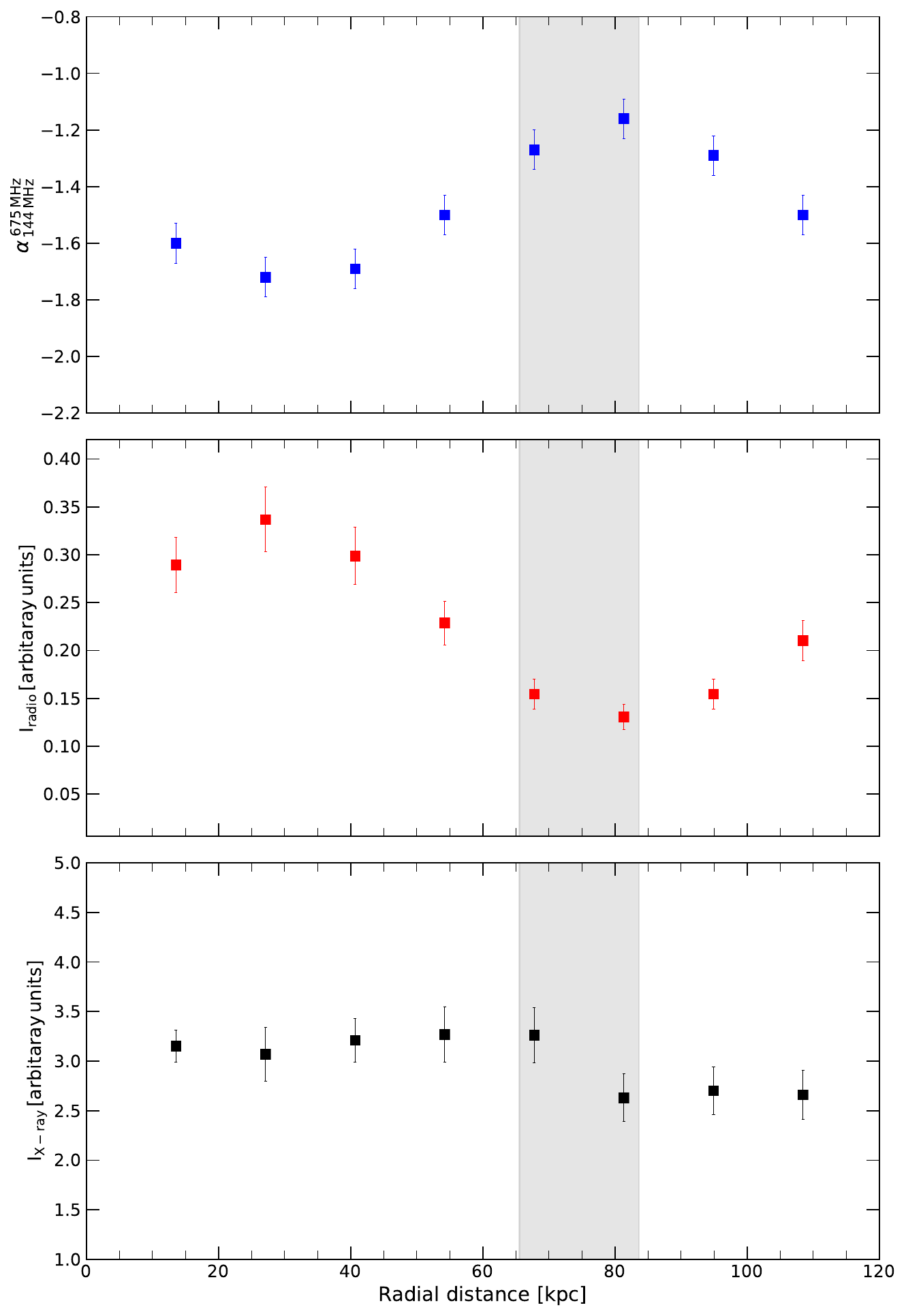}  
 \caption{Spectral index (Top), radio (Middle) and X-ray (Bottom) surface brightness profiles at the cold front CF3 (shown with gray shaded region). The width of each annulus is 12\arcsec.  The error bars represent both statistical and flux density scale errors. Both radio and X-ray profiles show a jump at the cold front.}
      \label{NW_radial_profiles}
\end{figure}

%%%%%%%%%%%%%%%%%%%%%%%%%%%
%Figure 15
%%%%%%%%%%%%%%%%%%%%%%%%%%%

 \begin{figure*}[!thbp]
    \centering
         \includegraphics[width=0.49\textwidth]{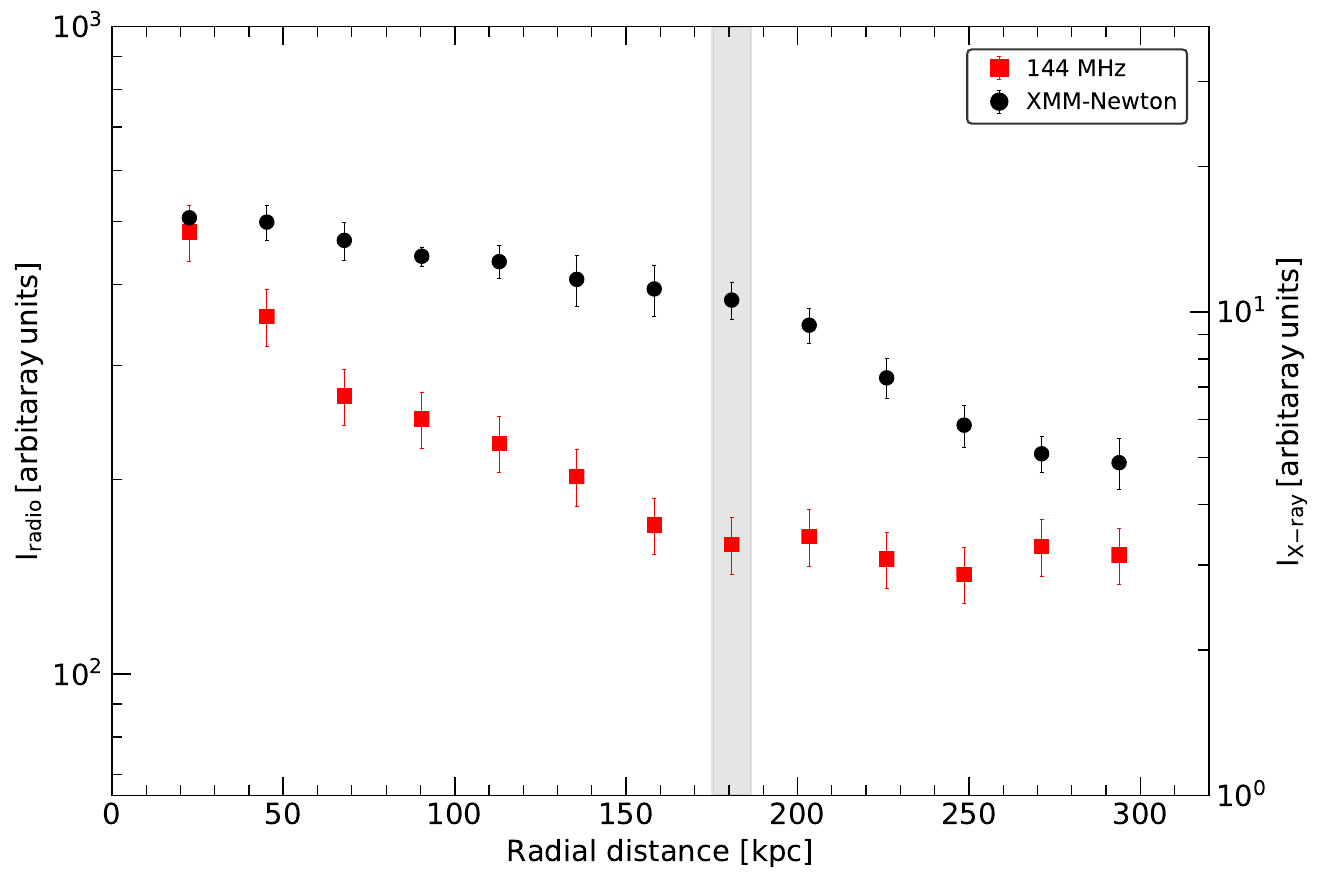}  
       \includegraphics[width=0.49\textwidth]{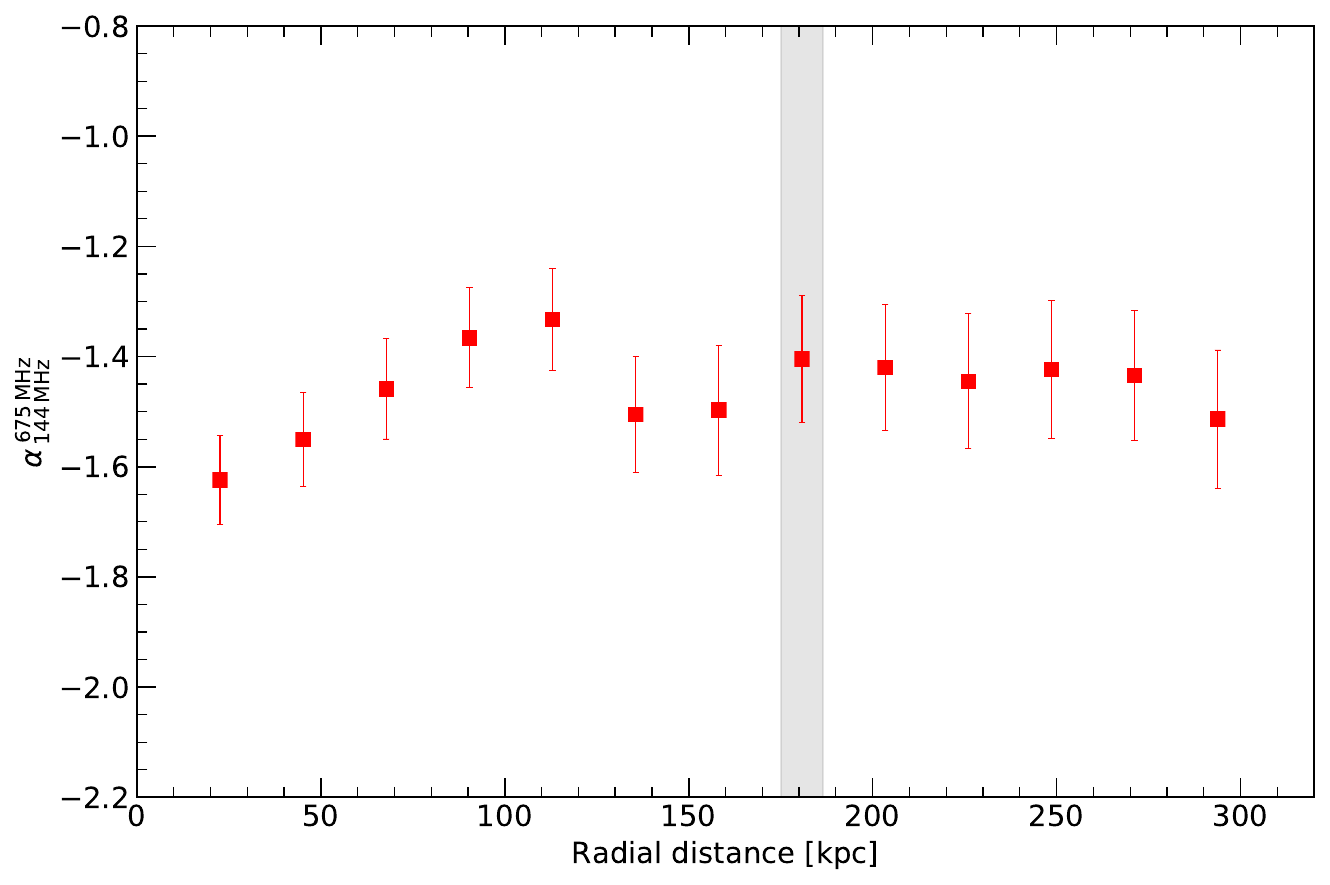}   
 \caption{\textit{Left}: Radio (Left) and spectral index (Right) profiles extracted in a sector at the cold front CF2 (shown with gray shaded region). Sector used for extracting the profile across CF2  is shown in Fig.\,\ref{subregions}. The width of each annulus is 20\arcsec. The error bars represent both statistical and flux density scale errors. Both radio and X-ray profiles show a jump at the cold front CF2. The sector used for extracting these profiles are shown in Fig.\,\ref{profile_regions}.  The gray shaded area for CF2 is smaller than the one shown in Fig.\,\ref{radial_profiles} because CF2 extends over various radii  (shown in red in Fig.\,\ref{profile_regions}) but in the current Figure the sector used for flux extraction is smaller (green sector in Fig.\,\ref{profile_regions}), i.e. smaller than the actual cold front length.}
      \label{SE_radial_profiles}
\end{figure*}

%############################################################################################################   
\subsection{Spectral index versus X-ray surface brightness}
%############################################################################################################   

We also checked the point-to-point relation between the spectral index and the X-ray surface brightness for the halo in Abell 2256. We divided the halo into three different regions: the core, innermost, and outermost regions (see Fig.\,\ref{IX_index} top panel). The data points associated with the wedge arc are also considered separately. To have high signal to noise ration, we only included regions where radio (at 144 and 675\,MHz) and X-ray surface brightness have emission above $3\sigma_{\rm rms}$. The resulting plot is shown in the bottom panel of Fig.\,\ref{IX_index}. There seems to be an anticorrelation between these two quantities: the fainter X-ray regions show the steepest spectral index. We again fit the data with a power law. The Spearman correlation coefficient is about $-0.60\pm0.10$. Therefore, there is a moderate anti-correlation between the spectral index and the X-ray surface brightness.  

Most of the X-ray emission is produced by the core in the main component of the cluster, where the radio spectral index is rather steep. It is plausible that the ICM and magnetic field in the core region has been moderately perturbed and stirred.  It is indeed possible that the dense core region is more shielded by gas dynamics during the merger and thus is less turbulent. This may explain why the spectral index measured in the core region is that steeper and also the fact that the overall point-to-point radio-to-X-ray surface brightness correlation is quite superlinear in the core region. If we exclude the core region of the cluster, the $\alpha-I_{\rm X}$ relation shows a strong correlation with the Spearman correlation coefficient of $-0.88\pm0.07$ (slope $b=0.09\pm0.01$), see bottom panel of Fig.\,\ref{IX_index}.

To our knowledge, the correlation between the spectral index and X-ray surface brightness has been studied so far only for four radio halos, with good statistical significance, namely in MACS\,J0717.5+3745, Abell 2255, Abell 2744, and ClG\,0217+70. In MACS\,J0717.5+3745 and Abell 2255 halos, a negative and positive correlation, respectively, have been found between these two quantities \citep{Rajpurohit2021b,Botteon2020a}. In contrast, for the halo in A2744, both positive and negative correlations have been reported, indicating a multicomponent halo \citep{Rajpurohit2021c}, while no significant correlation was found for the halo in CLG\,0217+70 \citep{Hoang2021}.

\subsection {Radial profiles} 
\label{radial_profile_section}
In order to investigate any connection between the radio emission and X-ray discontinuities, we created radial profiles using radio and X-ray maps. The mean of the radio and X-ray surface brightness and its errors were computed within concentric elliptical annuli, each having a width of 20\arcsec, i.e., the beam size of the radio maps. The annuli used are shown in Fig.\,\ref{profile_regions}. The area containing the relic and sources A, B, C, and F (magenta regions in Fig\,\ref{halo_index} bottom right panel) was not included in both radio and X-ray profiles. Moreover, other additional unrelated compact sources were also excluded; we simply masked them. We only included regions where the radio and X-ray surface brightness exceeds $3\sigma$. The resulting plot is shown in Fig.\,\ref{radial_profiles}. In this plot, the halo is about 0.8\,Mpc in diameter at 144\,MHz only.  The top panel shows the comparison of the radial profiles at 144 and 675\,MHz.  Radio radial profiles at these two frequencies follow more or less the same trend and decline sharply after 450\,kpc.  The innermost regions ($<450$\,kpc radii) are relatively flat. Typically, radio halos brightness profiles are described by an exponential law of the form 
\begin{equation}
I_{\rm r}=I_{0}e^{-\rm \frac{r}{r_{e}}},
\end{equation}    
where $I_{0}$ is the central radio brightness and $r_{\rm e}$ is the e-folding radius \citep{Murgia2009,Bonafede2022}. The halo in Abell 2256 shows a convex $\log(I)$ profile in the innermost regions (between 100-350\arcsec) instead of a concave profile, as observed for other known halos, e.g., Coma \citep{Bonafede2022}, Abell 2744 \citep{Orru2007,Murgia2009,Pearce2017}, and Abell 2255 \citep{Murgia2009}. Mini halos, that often host a central BCG, are known to show a convex profile in the innermost region \citep{Murgia2009,Chris2022}. We emphasize that the core region of the halo in Abell 2256 is apparently not associated with any clear radio galaxy.

The comparison of X-ray and radio profiles is shown in Fig.\,\ref{radial_profiles} middle and bottom panels. It is evident that the X-ray and radio profiles are different for a part of the halo emission. Radio profiles follow a convex curve while the X-ray is concave in the inner regions (<450\,kpc radii). After this, both the radio and the X-ray surface brightness drops rapidly to 800\,kpc. The 144\,MHz radio and X-ray profiles follow more or less the same trend from 450\,kpc to 800\,kpc. However, the 675\,MHz profile appears flatter than the X-ray profile. In Fig.\,\ref{radial_profiles} bottom panel, we show the radial profile of the dimensionless ratio ($\eta$) 
\begin{equation}
\eta=\frac{I_{\rm R}}{I_{\rm X}},
\end{equation}
between radio and X-ray surface brightness.

The halo in Abell 2256 does not show a uniform profile of $\eta$, see Fig.\,\ref{radial_profiles} bottom panel.  Both the GMRT and LOFAR profiles first show a decline up to a radius of 200\arcsec (i.e., 226\,kpc) followed by a sharp increase from 200\arcsec to 400\arcsec. The outermost regions (>400\arcsec) show a peculiar trend. It is worth highlighting that within 400\arcsec radii, two cold fronts are detected, shown with shaded areas in Fig.\,\ref{radial_profiles}. The shaded region is wider for CF2 because it extends over various radii. We note that the $\eta$ profile changes exactly at the location of the cold front CF2, also slightly at CF3, which implies a possible connection. The X-ray discontinuities may have an effect on the spectral and radio versus X-ray profiles. For the halo in Abell 2319, which shows both the halo and the mini-halo characteristic, a similar pattern was reported by \cite{Keshet_2010}: the profile first decreases in the innermost regions and then shows increasing trends. 

In Figs.\,\ref{NW_radial_profiles} and \ref{SE_radial_profiles}, we show the radio and X-ray profiles extracted separately at the cold fronts CF3 and CF2. The sectors used for extracting these profiles are shown in Fig.\,\ref{profile_regions}. For the innermost cold front CF3, there is a jump in both radio (middle) and X-ray (bottom) surface brightness: both decreases at the cold front. A jump is also visible in the spectral index profile (top panel); the spectrum becomes flatter at the cold front. 

Similarly, CF2 also shows a small jump in the radio surface brightness profile, see Fig. \,\ref{SE_radial_profiles}. In contrast to CF3, here the radio and X-ray profiles are quite different for radius $<150\rm \,kpc$;  characterized by convex (radio) and concave (X-ray) shapes. At the cold front CF2, the X-ray brightness decreases, whereas the radio brightness increases. Unlike CF3, the spectral index profile for CF2 does not show a clear jump.

\subsubsection{Origin of the Core}
The core has a steep spectral index and its radio vesus X-ray brightness correlation is superlinear. Furthermore, it is confined by the innermost cold front (CF3). A sublinear slope in the radio versus X-ray relation is expected for mini halos (relaxed clusters) produced by sloshing motions \citep{Ignesti2020}. Origins for the radio halo core could be:
\begin{enumerate}
\item{} Old radio plasma from previous AGN activity (similar to mini halos) is advected, compressed, and  reaccelerated by the activity of the cold front CF3. Simulations indeed suggest that CRe and magnetic field from former activity, confined by cold fronts, can be mixed and advected to the ICM \citep{ZuHone2021}. In this case, it is difficult for CRe to travel from the inside across the cold front to the outside because the magnetic field is probably stirred and aligned along the cold front \citep{ZuHone2011}. We note that ultra-low frequency radio observations show that the spectral index flattens in the core below 43\,MHz (Osinga et al. in prep). The flattening at frequency below 43\,MHz seems consistent with the fact that an old plasma mixed and reaccelerated in a relatively small volume (quasi-homogeneous conditions).

\item{} Turbulence is less strong in the core with respect to the surrounding region, and the magnetic field is significantly higher than $B_{\rm cmb}/\sqrt{3}$. Under these conditions, the radio emissivity could be high but the radio spectrum is steep. 
\end{enumerate}

\subsubsection{Emission outside the halo core}

The radio X-ray correlation slope steepens when moving from region1 to region2, followed by a flatter slope in region3  while the spectral index steepens radially from region1 to region3. The cold front CF2 lies within region2. Furthermore, the southern shock front falls within region3, where the correlation is almost negligible, as expected in the cluster outskirts for radio relics. However, we have not detected any spectral index gradient associated with the southern shock front. The change in the correlation slopes and spectral indices from region1- region3 could be due to:
\begin{enumerate}
\item Radial decline of magnetic field : as soon as $B \leq B_{\rm cmb}/\sqrt{3}$,  the radio radial profile becomes steeper, not necessarily superlinear with X-rays (Fig.\,\ref{radial_profiles}).
\item  Increase in the number density of seed particles with radius: less effective Coulomb losses (lower thermal density) and the accumulation of light plasma that is buoyantly transported to the outskirts may increase the number density of available seed electrons and consequently increase the radio emissivity. This results in flattening of the radio versus X-ray relations. Therefore, the flattening of the radio vs X-ray correlation slopes from region2 to region3 (Fig.\,\ref{subregions}) could be due to an increase in the number density of seed particles. We emphasize that this does not necessarily produce a flattening in the integrated radio spectrum (Fig.\,\ref{halo_index}) as well. 
\item  More turbulence in the external regions: this results in decrease in $\tau_{\rm acc}$, i.e., increases efficiency. As a result, the radio emissivity will increase and the radio spectrum will flatten (for a fixed magnetic field value). This flattening may be balanced by a decline of magnetic field. Since, we do not detect any spectral flattening from region1 to region3, this possibly suggests more radial magnetic field radial decline). 
\end{enumerate}

%############################################################################################################   
\subsection{Wedge arc}
In our total power images (Fig.\,\ref{low_res}), the wedge arc appears to connect the large relic, the halo and source F. The integrated spectral index of the wedge arc between 144\,MHz and 1.5\,GHz is $\alpha=-1.63\pm0.05$ (see Fig.\,\ref{halo_index}). This value is very similar to the spectral index of the halo region. 
%############################################################################################################   

The arc is near the outer edge of the detected \textit{Chandra} X-ray emission. However, no X-ray discontinuity was reported at that location. We do not find any spectral index gradient in this region. Polarized emission is not detected in the wedge arc region \citep{Owen2014}. Furthermore, compared to the larger relic ($\alpha=-1.07$), the  arc shows a much steeper spectral index, similar to that of the halo. This, at least rules, out the relic interpretation. It cannot be a radio phoenix, possibly originating from source F because no curvature is found in the overall spectrum.  In the radio spectral index versus X-ray surface brightness correlation, the data points from this region show a trend very similar to that found in the radio halo.

One possibility of explaining the wedge arc is that the relativistic plasma from the halo is advected by a complex patterns of fluid motions (vortex-like) that is generated in the tail of the shock of the radio relic. If the spatial diffusion of particles and magnetic field is not fast, relativistic particles and magnetic field are attached to the thermal plasma and basically follow the velocity flow of the gas. The passage of a subcluster (and shock) may thus generate eddies/vortex at the boundaries downstream that may ``shape'' the boundary of the halo similar to an arc-shaped wedge structure. In conclusion, on the basis of similar spectral indices and similar trends in the correlation of radio spectral index versus X-ray surface brightness, the radio emission in the wedge arc is most likely associated with the halo, which could be seen in projection with the relic and source F. 

%############################################################################################################   

\subsection{Source AI}
%############################################################################################################   
Source AI is located in the outskirts of the cluster and shows an ultra steep spectral index, about $-1.9$ between 350 and 675 MHz. The source is not detected at 1-2\,GHz and the integrated spectral index between 144\,MHz and 675\,MHz is already curved. Optical images do not reveal any obvious counterpart \citep{vanWeeren2009a}. The source could be a AGN remnant or radio phoenix. AGN remnants originate from a previous episode that left a fossil population from a nearby AGN. On the other hand, radio phoenices are produced by adiabatic compression of AGN fossil radio plasma by the passage of a shock front in the ICM \citep{Ensslin1998}. 

In the high resolution radio images, we cannot identify any related radio galaxy associated with AI. On the basis of the morphology, location, total extent, steep spectral index, and curved integrated spectrum, we classify the source as a radio phoenix.

%############################################################################################################   
\section{Summary and conclusions}
\label{summary}
%############################################################################################################   
In this work, we have presented a new detailed analysis of the radio halo in the famous galaxy cluster Abell 2256. Previous studies of the halo have been limited by their low resolution and sparser \textit{uv}-coverage. Our deep , high fidelity observations, in combination with the available X-ray observations, provide crucial insights into the origin of the radio halo. We summarize our main findings as follows:

\begin{enumerate} 
\item We show the first deep, high spatial resolution (12-20\arcsec) radio images of the central halo. The halo emission is detected at all frequencies, namely 144\,MHz, 350\,MHz, 675\,MHz and 1.5\,GHz. The LLS size of the halo is about 0.9 Mpc at 144 MHz and 0.5\,Mpc at 1.5 GHz, implying that the outermost region of the halo has a steep spectrum.\\

\item Despite the complex ICM distribution, the radio halo morphology is remarkably similar to that in X-ray. In particular, the X-ray peak in the main mass component coincides with the radio peak. Moreover, the radio surface brightness, within the main component, is confined by the innermost cold front. \\

\item The overall emission from the halo follows a power-law spectrum between 144\,MHz and 1.5\,GHz, and has an ultra steep spectrum with an integrated spectral index of $-1.63\pm0.03$. According to turbulent reacceleration models,  the absence of any steepening at high frequency in the integrated spectrum suggests that the emitting volume is not homogeneous. At the cluster core (or centroid), we find evidence of an ultra steep spectral index ($\alpha=-1.60$) while the surrounding region is flatter ($\alpha=-1.45$).\\

\item  The spatially resolved spectral index maps reveal a spectral steepening with increasing radius. We  find small-scale fluctuations in the spectral index across the halo, on scales of about 20\,kpc.\\

\item The halo fits well with the known radio power-mass correlation of other halos at 144\,MHz. However, it is significantly underluminous with respect to the radio power-mass correlation at 1.4\,GHz. This is in line with current models where ultra steep spectrum are generally underluminous. \\

\item The point-to-point comparison between the radio and X-ray surface brightness across the halo reveals a strong sublinear correlation. The correlation slope flattens at higher frequencies. There are substructures: the halo core shows a superlinear slope ($\rm I_{R}\propto I_{X}^{1.51}$)  while the outermost fainter region shows a sublinear slope ($\rm I_{R}\propto I_{X}^{0.41}$). We suggest the core emission is related to old plasma from previous AGN activity, being advected, compressed and re-accelerated by the nearby cold front. Another possibility  is that the turbulence is less strong in the core and magnetic field is high making the spectrum steep. The change in the radio vs X-ray correlation slope in the other region2 and region3 could be due to magnetic field radial decline, increasing number density of seed particles or more turbulence in the outermost regions. \\

\item We find a strong anti-correlation between the spectral index and the X-ray surface brightness across the halo. This is consistent with radial steepening, the outermost steeper part corresponds to regions with less dense ICM, i.e., less turbulence and a decline in the magnetic fields.\\

\item The  radio spectral index versus X-ray correlation across the arc-shaped wedge structure to the east of the cluster center, connecting the source F to the relic and the halo, show  trends similar to that of the halo. The integrated spectrum of this structure is also very similar to the halo, hinting that the radio emission in the wedge arc  is likely a part of the halo. \\  

\item  A diffuse elongated source to the southeast (source AI) fits the profile of a radio `phoenix' produced by adiabatic compression of fossil radio plasma due to a merger shock front.

\end{enumerate} 
To summarize, our new results stress once more that the radio halo in Abell 2256 is a very peculiar object, whose study can lead to new progress in our understanding on particle acceleration mechanisms on very large scales. First, a number of observed features are is in line with turbulent re-acceleration models for the origin of radio emitting electrons. For example, its ultra steep integrated spectrum, the different spectral indices in the subregions, the spectral steepening with increasing radius, the spectral index fluctuations, the sublinear radio versus X-ray correlation slope and the fact that it is underluminous at high frequency. On the other hand, not all the observable evidence can be explained by turbulent acceleration models, requiring for updates on the model (e.g., the ultra steep spectral index and the superlinear radio versus X-ray correlation slope in the core region, the connection of the radio emission with the cold fronts, and the radio versus X-ray correlation trends in the halo subregions). Our results suggest that none of the  proposed models for halo formation can explain all the observational findings at the same time. 

In conclusion, the emerging complexity in the radio halo properties that we are able to observe with the new generation radio telescopes is providing clear evidence that homogeneous models are too simple to explain all the observed features in this class of sources. For example, the Coma \citep{Bonafede2022} and Abell 2256 halos are among the closest clusters, and we already started seeing differences between them. Our results highlight the need for additional theoretical work, with  more detailed and specific predictions, on particle acceleration in radio halos.

%####################################
\begin{acknowledgements}
%####################################

KR and FV acknowledge financial support from the ERC Starting Grant ``MAGCOW'' no. 714196. RJvW acknowledges support from the ERC Starting Grant ClusterWeb 804208. AB acknowledges support from the VIDI research programme with project number 639.042.729, which is financed by the Netherlands Organisation for Scientific Research (NWO). A. Botteon, A. Bonafede, CJR, EB, and CS acknowledges support from the ERC through the grant ERC-Stg DRANOEL n. 714245. W.F. acknowledges support from the Smithsonian Institution and the Chandra High Resolution Camera Project through NASA contract NAS8-03060. AD acknowledges support by the BMBF Verbundforschung under the grant 05A20STA. MB acknowledges support from the Deutsche Forschungsgemeinschaft under Germany's Excellence Strategy-EXC 2121 ``Quantum Universe" -390833306. MR acknowledges support from INAF mainstream project ``Galaxy Clusters science with LOFAR''. The GMRT is run by the National Centre for Radio Astrophysics (NCRA) of the Tata Institute of Fundamental Research (TIFR). 
LOFAR \citep{Haarlem2013} is the Low Frequency Array designed and constructed by ASTRON. It has observing, data processing, and data storage facilities in several countries, which are owned by various parties (each with their own funding sources), and that are collectively operated by the ILT foundation under a joint scientific policy. The ILT resources have benefited from the following recent major funding sources: CNRS-INSU, Observatoire de Paris and Universit\'{e} d'Orl\'{e}ans, France; BMBF, MIWF-NRW, MPG, Germany; Science Foundation Ireland (SFI), Department of Business, Enterprise and Innovation (DBEI), Ireland; NWO, The Netherlands; The Science and Technology Facilities Council, UK; Ministry of Science and Higher Education, Poland; The Istituto Nazionale di Astrofisica (INAF), Italy. This research made use of the LOFAR-UK computing facility located at the University of Hertfordshire and supported by STFC [ST/P000096/1], and of the LOFAR-IT computing infrastructure supported and operated by INAF, and by the Physics Dept. of Turin University (under the agreement with Consorzio Interuniversitario per la Fisica Spaziale) at the C3S Supercomputing Centre, Italy.  
\end{acknowledgements}

\bibliographystyle{aa}

\bibliography{ref.bib}

\end{document}